\documentclass[article]{JHEP3} % 10pt is ignored!

\usepackage{epsfig,multicol,bbm, graphicx}
\usepackage{amsmath}

\newcommand{\eq}{\begin{equation}}
\newcommand{\eqx}{\end{equation}}
\newcommand{\eqn}{\begin{eqnarray}}
\newcommand{\eqnx}{\end{eqnarray}}
\newcommand{\N}{{\cal N}}

\newcommand{\nn}{\nonumber \\}

\newcommand{\f}[2]{\frac{#1}{#2}}
\newcommand{\cor}[1]{\left\langle{#1}\right\rangle}
\newcommand{\eps}{\varepsilon}

\title{Boost-invariant early time dynamics from AdS/CFT}

\date{23.06.2009}
\maketitle
\author{Guillaume Beuf$\,^a$, Michal P. Heller$^b$, Romuald A. Janik$^b$ and Robi Peschanski$^a$
\footnote{{\tt guillaume.beuf@cea.fr, michal.heller@uj.edu.pl, ufrjanik@th.if.uj.edu.pl, robi.peschanski@cea.fr}}
\\ \\ \\ 
\small{\emph{$^{a}$ Institut de Physique Th{\'e}orique URA 2306, Unit\'{e} de Recherche associ{\'e}e au CNRS,}} \\
\small{\emph{CEA-Saclay, F-91191 Gif/Yvette Cedex, France.}} \\ \\
\small{\emph{$^{b}$ M.Smoluchowski Institute of Physics, Jagellonian University,}} \\
\small{\emph{Reymonta 4, 30-059 Krakow, Poland.}} }

\abstract{Boost-invariant dynamics of a strongly-coupled  conformal plasma
is studied in the regime of early proper-time using the AdS/CFT correspondence.
It is shown, in contrast with the late-time expansion,  that a scaling solution
does not  exist. The boundary dynamics in this regime depends on initial
conditions encoded in the bulk behavior of a Fefferman-Graham metric
coefficient at initial proper-time. The relation between the early-time 
expansion of the energy density and initial conditions in the bulk of AdS is
provided. As a general result it is proven that a singularity of some metric
coefficient in Fefferman-Graham frame exists at all times. Requiring that this
singularity at $\tau = 0$ is a mere coordinate singularity without the curvature
blow-up gives constraints on the possible boundary dynamics. Using a simple Pade
resummation for  solutions satisfying the regularity constraint, the features
of a transition to local equilibrium, and thus to the hydrodynamical late-time
regime, have been observed. The impact of this study on the problem of
thermalization is discussed.}

\keywords{Gauge-gravity correspondence, Heavy Ions}

\begin{document} 

%\maketitle  IS IGNORED %%%%%%%%%%%

\section{Introduction}

Recently, a new theoretical approach to heavy-ion reactions and the formation of
quark-gluon plasma (QGP) in RHIC experiments using the AdS/CFT correspondence
arose as a fruitful application of string theory to the real world.
The AdS/CFT correspondence allows to describe many properties of a class
of gauge theories at strong coupling in terms of a dual gravitational
description. What makes it a unique tool is firstly that it works well in
Minkowski signature allowing for a description of inherently time-dependent
phenomena, in contrast to lattice QCD methods which are directly tied to
Euclidean signature.
Secondly, explicit analytical computations are often possible and one can get
new insight into strongly coupled dynamics from the dual geometric perspective.

On the experimental side, the features of the distribution of particles 
observed in the final stage of the reaction point to the existence of a
hydrodynamical regime of the QGP expansion \cite{review,hydro}, 
in particular the observed  elliptic flow denoting  sizeable collective
effects \cite{JY}. The hydrodynamic regime has to last long enough and 
start soon enough after the collision in order to explain the observed
collective effects. Moreover, the smallness of
the viscosity which can be extracted from hydrodynamical simulations describing
the data leads to an almost-perfect fluid behaviour of the QGP, and thus to a
short mean-free path inside the fluid. Putting together these experimental
inputs, and in order to go beyond a mererly phenomenological description, it
appears to be theoretically necessary to investigate as much as possible the
properties of a strongly-coupled Quantum-Chromodynamic plasma.

In the absence of nonperturbative methods applicable to real-time
dynamics of strongly coupled Quantum Chromodynamic (QCD) plasma, one is led 
to consider similar problems from the point-of-view of the AdS/CFT
correspondence, that is looking for the characteristics of plasma
in a gauge theory for which the AdS/CFT correspondence takes its simplest form
-- the $\N=4$ supersymmetric Yang-Mills theory \cite{adscft} which possesses
a known and tractable gravity dual.

Although the $\N=4$ gauge theory is supersymmetric and conformal and thus
quite different from QCD at zero temperature, both supersymmetry and
scale-invariance are broken explicitly at finite temperature and we
may expect qualitative similarities with QCD plasma for a range of temperatures
above the QCD deconfinement phase transition\footnote{There exist more refined
versions of the AdS/CFT correspondence which may have more features in common
with QCD, however the gravity backgrounds are much more complicated and we will
not consider them here.}.

Indeed, the gauge/gravity dual calculation \cite{son} showing, in a static
setting, that the viscosity over entropy ratio $\eta /s$ is very small (equal
to $1/4\pi$) and
even suggesting a universal lower bound, is in qualitative agreement 
with hydrodynamic simulations of QCD plasma and was a poweful incentive to
explore further the AdS/CFT duality approach.

\begin{figure}
    \begin{center}
    \scalebox{0.9}[0.9]{\includegraphics[]{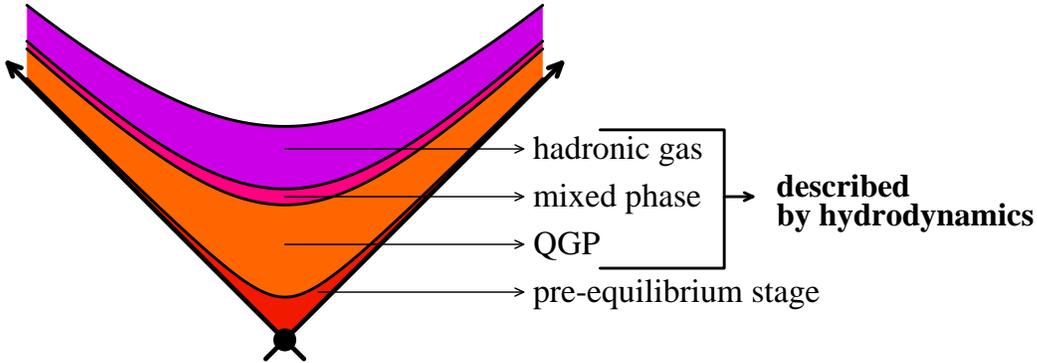}}
    \caption{Description of QGP formation in heavy ion collisions. The 
kinematic landscape is defined by ${\tau = \sqrt{(x^0)^2-(x^3)^2}\ ;\ {y=\f 12 
\log \f 
{x^0+x^3}{x^0-x^3}}\ ;\ {x_\bot\!=\!\{x^1,x^2\}}}\ ,$
where the coordinates along the light-cone are $x^0 \pm x^1,$ the transverse 
ones 
are $\{x^1,x^2\}$ and $\tau$ is the proper time, $y$ the ``space-time 
rapidity''.}
    \end{center}
\label{kinematics}
\end{figure}
In order to go beyond static calculations, one has to adapt the dual 
AdS/CFT approach to  the relativistic kinematic framework of heavy-ion
reactions, where two ultra-relativistic heavy nuclei collide and form an
expanding medium, see Fig.\ref{kinematics}. It is convenient, initially, to
adopt the simplest setting in which the relevant physics is still present.

For this sake, one is  guided by the seminal work of Bjorken \cite{Bjorken}, who
made a simplifying assumption of boost-invariant dynamics for his hydrodynamic
description of high-energy heavy-ion collisions. Hence, the different stages of
the plasma formation, expansion and hadronization, see Fig.\ref{kinematics}, 
depend only on the proper-time $\tau$ and not on rapidity\footnote{For the
Bjorken flow, spatial and momentum rapidity are equal.} $y.$ This assumption is
indeed phenomenologically valid for the central rapidity  region of the
collision at high energy. 

For the application of the AdS/CFT correspondence to this setting, and in
particular for finding the gravity dual \cite{US1}, the above symmetry
assumptions allow for a drastic simplification of an otherwise yet probably
intractable problem, since the dual dynamics depends only on two variables,
$\tau,$ proper-time and $z,$ the bulk variable in the dual asymptotically AdS
spacetime. 

The two main physical questions\footnote{We do not address here  another
important  problem which is hadronization, since it requires to go beyond our
current understanding of the
presently known Gauge/Gravity dualities.} are to understand the behaviour of
quark gluon plasma at late and early times. 
At late times, we would like to {\em derive} the appearance of hydrodynamic
behaviour with all its characteristics,  asymptotically perfect fluid flow, low
viscosity, transport coefficients, etc.
At early times, we would like to understand the rapid thermalization starting
from the ultra-relativistic initial conditions. 

In fact the study of the former problem has led to quite successful
developments,
showing that the hydrodynamic boost-invariant Bjorken flow is a consequence of
late-time dynamics of gravity in the dual side of the AdS/CFT correspondence.
Subsequently the way in which
hydrodynamics arises from the AdS/CFT has been understood even without using any
symmetry arguments \cite{MINWALLA} as a gradient expansion of Einstein's
equations. However since our main interest in this paper is the radically
different early time regime where such expansions are inapplicable, we will use
the full system of Einstein's equations with the boost-invariant symmetry assumptions.
Before that, we will first recall the analysis for late times, since the point of
departure will be analogous for the early time problem, but with different
developments and results. 

The second problem, which forms the novel part of our investigations in the 
present study, is to go beyond hydrodynamics and investigate early-time
dynamics. The only assumption we will make, apart from boost-invariance, is the
existence of a strongly coupled pre-hydrodynamic stage\footnote{Physically, it
may be that the initial stage of the collision has some weak-coupling
properties, for instance in the Color Glass Condensate (CGC) setting \cite{CGC}, but
we will not consider here this possibility (for a recent attempt to describe CGC 
initial conditions within the AdS/CFT see \cite{beuf}).}
allowing for the use of the AdS/CFT correspondence in the same framework as for
late times, and thus to look for a connection of the two regimes in a unified
scheme. In particular, one is curious to understand the transition towards the
hydrodynamic regime and thus to get some information about the puzzling problem of
the apparently short \cite{hydro} thermalization time.

Summarizing the contents of the paper, the next section recalls the basic
AdS/CFT approach to boost-invariant dynamics. The main results of the
late-time solution are given in section 3, with emphasis on those aspects which
will be useful for the early-time regime. Section 4 is devoted to the
theoretical analysis of the early-time boost-invariant dual dynamics, while
section 5 deals with numerical solutions with the goal of studying the
transition to hydrodynamics.  Conclusions and outlook are given in the final
section.

\section{Boost-invariant dynamics and holography}
\subsection{Energy-momentum tensor of boost-invariant flow}

Currently the best understanding of non-perturbative non-equillibrium
4-dimensional quantum field theories is provided by string theory 
methods. Holography translates gauge theory problems in four dimensions at
large number of colors and strong ('t Hooft) coupling 
into higher-dimensional classical gravity. Einstein's equations involve then, in
the simplest setting, functions of both gauge theory 
dimensions and of one more variable $z$ parameterizing the radial direction in
AdS. In a general setup, finding solutions of Einstein's equations must
involve 
numerical analysis and is rather difficult. Moreover, singular components of the
metric may and indeed do appear 
\cite{US1,RJ,Heller:2007qt} which are not easy to implement numerically, unless
their structure is known beforehand. It is thus 
important to focus on the simplest possible, yet phenomenologically 
interesting, gauge theory dynamical processes. As explained in the
introduction, we will focus on the boost-invariant plasma
expansion. This setup serves as a first approximation of dynamical plasma
created at RHIC, as excited nuclear 
matter undergoes one-dimensional expansion and is both translationally and
rotationally invariant in the plane perpendicular to the 
expansion axis (physical quantities do not depend on $x^{1},x^{2}$). The
crucial simplifying assumption is boost-invariance along the expansion axis,
which can be most easily imposed when passing to  proper-time $(\tau)$ and
(spatial) rapidity 
$(y)$ coordinates related to the usual lab-frame time $(x^{0})$ and position 
along the expansion axis $(x^{3})$ by 

\begin{eqnarray}
x^{0} &=& \tau \cosh{y} \nonumber \\
x^{3} &=& \tau \sinh{y}
\ .
\end{eqnarray}
Under a boost along the $x^{3}$ direction proper-time does not change, whereas 
rapidity is shifted by a constant. Thus in this coordinate frame
boost-invariance means that physical observables do not depend on rapidity. 

A quantity of special interest is the expectation value of the energy-momentum
tensor of the plasma. In the $(\tau,y,x^{1},x^{2})$ frame with all 
mentioned symmetries\footnote{Together with parity invariance in rapidity,
which ensures that there are no nondiagonal $T_{\tau\, y}$ 
terms in the energy-momentum tensor.} imposed, the most general boost-invariant
energy-momentum tensor takes the form

\begin{equation}
\label{e.tmunudiag}
T_{\mu \nu} = \mathrm{diag} \left\{ \epsilon\left( \tau \right), \, \tau^{2}
p_{\parallel} \left( \tau \right),\, p_{\perp} \left( \tau 
\right),\, p_{\perp} \left( \tau \right) \right\}\ \mathrm{.}
\end{equation}
Imposing energy-momentum conservation $\nabla_{\mu} T^{\mu \nu} = 0$ (in
boost-invariant case there is only one independent equation) and tracelessness 
$T^{\mu}_{\mu} = 0$ leads to the energy momentum tensor fully specified in
terms of a single function of one variable -- the energy 
density $\epsilon\left( \tau \right),$ namely

\begin{equation}
T_{\mu \nu} = \mathrm{diag}  \left\{ \epsilon\left( \tau \right), \, - \tau^{2}(
\epsilon \left( \tau \right) + \tau \epsilon ' \left( 
\tau \right)), \, \epsilon  \left( \tau \right) + \frac{1}{2} \tau \epsilon '
\left( \tau \right) ,\, \epsilon \left( \tau \right) + 
\frac{1}{2} \tau \epsilon ' \left( \tau \right) \right\} \mathrm{.}
\end{equation}
The function  $\epsilon\left( \tau \right)$ may be arbitrary at this level and
it is the gravitational dual which will single out the 
allowed ones.
Note that  $T_{\mu \nu}$ is submitted to an important positivity constraint,
namely $T_{\mu\nu}t^\mu t^\nu \ge 0,$ for any time-like vector $t^\mu$ implying
\cite{US1}

\begin{equation}
\label{positivity}
- \f{4\ \epsilon\left( \tau \right)}\tau\ \le\ \epsilon'\left( \tau \right)\
\le\ 0  \mathrm{.}
\end{equation}
This  requirement may perhaps be lifted in case of some highly quantum states
\cite{REF} appearing at transient times, but we expect that it should hold for any
more macroscopic configuration. We observe a temporary violation of this
constraint for a specific initial condition (see section 5).

\subsection{Holographic reconstruction of space-time}

The AdS/CFT correspondence maps the expectation value of the energy-momentum
tensor operator into the behavior of the 5-dimensional 
asymptotically AdS metric $G_{AB}$. This metric is a solution of Einstein's
equations with negative cosmological constant $\Lambda = - 6 
\, L^{-2}$ (where L is the AdS radius\footnote{In the rest of the paper the AdS
radius L will be set to 1, but can be trivially restored on 
dimensional grounds. Greek and latin indices denote respectively field theory
and bulk AdS directions.}), namely

\begin{equation}
\label{EINSTEINeqns1}
R_{AB} - \frac{1}{2} R \cdot G_{AB} - 6 \, G_{AB} = 0\mathrm{.}
\end{equation}
The general asymptotically AdS metric written in  Fefferman - Graham
coordinates \cite{fg} takes the form

\begin{equation}
\mathrm{d} s^{2} = G_{A B} \mathrm{d}y^{A} \mathrm{d}y^{B} = \frac{g_{\mu \nu} \mathrm{d} x^{\mu} \mathrm{d} x^{\nu}  + \mathrm{d} 
z^{2}}{z^{2}}
\end{equation}
where $z$ is the AdS radial variable and $g_{\mu \nu}$ is a four-dimensional
metric on constant $z$ slices, which depends both on the 
field theory directions $x^{\mu}$ and $z$. The hypersurface at $z=0$ is a
conformal boundary of AdS and is identified with the metric 
in which the CFT lives. In the considered field theory setup this is the
4-dimensional Minkowski metric $\eta_{\mu \nu}$.  Einstein's
equations obtained from \eqref{EINSTEINeqns1}) (see further Eq. \eqref{einsteinequations}
for the full set of Einstein's equations 
expressed in Fefferman-Graham coordinates) can be solved pertubatively in the
vicinity of the conformal theory. The four-dimensional 
part $g_{\mu \nu}$ of the AdS metric\footnote{All the formulas regarding
holographic reconstruction of the bulk space time are provided 
for the case when the boundary is the Minkowski space.} 

\begin{equation}
\label{metricEXP}
g_{\mu \nu} = g_{\mu \nu}^{\left( 0 \right)} + g_{\mu \nu}^{\left( 2 \right)}
\, z^{2} + g_{\mu \nu}^{\left( 4 \right)} \, z^{4} + 
g_{\mu \nu}^{\left( 6 \right)} \, z^{6} + + g_{\mu \nu}^{\left( 8 \right)} \,
z^{8} + \ldots
\end{equation}
can be obtained order by order in $z^{2}$ from the equations of motion
(\ref{EINSTEINeqns1}). The leading piece in (\ref{metricEXP}) is 
the Minkowski metric $\eta_{\mu \nu}$, which in the proper $(\tau,y,{\textbf x_\bot})$
variables takes the form

\begin{equation}
g^{\left(0\right)}_{\mu \nu}  \mathrm{d} x^{\mu} \mathrm{d} x^{\nu} = \eta_{\mu
\nu}  \mathrm{d} x^{\mu} \mathrm{d} x^{\nu}  = - 
\mathrm{d} \tau^{2} + \tau^{2}  \mathrm{d} y^{2} + \mathrm{d}
\mathrm{\textbf{x}}_{\perp}^{2} 
\ .
\end{equation}
The quadratic term $g_{\mu \nu}^{\left( 2 \right)}$ vanishes identically, and
the first non-trivial term $g_{\mu \nu}^{\left( 4 
\right)}$ turns out to be proportional to the expectation
value of the energy-momentum tensor \cite{Skenderis}
\begin{equation}
g_{\mu \nu}^{\left( 4 \right)} =  \frac{2 \pi^{2}}{N_{c}^{2}} \cor{T_{\mu \nu}}
\mathrm{.}
\end{equation}
Higher order contributions ($g_{\mu \nu}^{\left( 2 n\right)}$ for $n > 2$) can
be obtained from the Einstein's equations and are fully 
expressed in terms of $\cor{ T_{\mu \nu} }$ and its derivatives. At this level
there are no constraints on the form of the 
energy-momentum tensor -- any conserved and traceless one will do (apart from
the positivity conditions \eqref{positivity}). 

However it is to be expected \cite{US1} that generic energy-momentum tensor
would lead to singularities in the bulk of AdS, which are 
beyond the scope of a perturbative solution in $z^{2}$. Therefore in order to
have a control over the geometry in the region where 
potential singularities may develop, one needs to solve the Einstein's equations
deeper in the bulk. Since the energy-momentum tensor for 
the boost-invariant flow takes a particularly simple diagonal form
(\ref{e.tmunudiag}), in this 
case we are led to adopt the 5-dimensional Fefferman-Graham metric ansatz
satisfying the symmetries of boundary dynamics

\begin{equation}
\label{metricANSATZ}
\mathrm{d} s^{2} = \frac{ - e^{a\left( \tau, z \right)} \mathrm{d} \tau^{2} + \tau^{2} e^{b\left( \tau, z \right)} \mathrm{d} y^{2} + 
e^{c \left( \tau, z \right)} \mathrm{d} \mathrm{\textbf{x}}_{\perp}^{2}+ \mathrm{d} z^{2}}{z^{2}} \mathrm{.}
\end{equation}
The full set of Einstein's equations corresponding to the parametrized metric
\eqref{metricANSATZ} are given in \eqref{einsteinequations}. 
Solving the Einstein's equations
for any value of $\tau$ and $z$ is very difficult and must 
probably involve numerical procedures. However it is possible to gain
nevertheless some analytical insight.
There are two regimes where 
some simplifications may occur -- these are for small or large values of $\tau$.
In both cases there is a small parameter (either $\tau$ or 
its inverse), which can be used to construct the geometry in perturbative
manner in $\tau$, but exactly in $z$. So we shall focus on late 
and early time dynamics.

\section{Late time dynamics}
\subsection{Motivation}

At sufficiently late times, the boost-invariant plasma is expected to be 
locally equilibrated and thus well-described by 
hydrodynamics. However, it is {\it a priori} not obvious how to prove it 
on the gravity side of the AdS/CFT. The solution was 
first shown in \cite{US1} and then further
developed in \cite{Nak1, RJ, Bak:2006dn, Heller:2007qt, 
Benincasa:2007tp, Buchel:2008xr, Heller:2008mb, Kinoshita:2008dq} (see
\cite{Heller:2008fg} for a review). The following subsections 
review how to obtain the late-time boost-invariant dynamics perturbatively,
since the methods developed for this
approach will be crucial in the rest of the paper when considering the open
problem of early time dynamics and the corresponding 
transition to local equilibrium.

\subsection{The asymptotic solution}

The goal of the whole approach both in the early and late times domain is to
solve the Einstein's equations \eqref{einsteinequations} 
perturbatively in proper time $\tau$, but (at each order in $\tau$) exactly in
the radial AdS variable $z$. This can be achieved by 
resumming the power series in $z^{2}$ for the bulk metric (\ref{metricEXP}).
For arbitrary proper-time dependence of the energy density it is yet out of reach, but 
focusing on the large proper time regime and assuming that the energy density
of the boost-invariant plasma exhibits a power-like tail

\begin{equation}
\epsilon\left( \tau \right) \sim \f \# {\tau^{s}} + \ ...
\end{equation}
for some positive power\footnote{The power $s$ is constrained to take values in
the range $0 \leq s < 4$ assuming positivity of the energy 
density \eqref{positivity} in any time-like reference frame. There are three
particular phenomenologically interesting values of $s$ 
within this range: $s = 4/3$, $s = 1$ or $s=0$. The first  one leads to the {\it
perfect fluid} case where $p_{\perp} \left( \tau \right) = 
p_{\parallel} \left(\tau \right)$ and the second one to the {\it free streaming}
scenario with $p_{\parallel} \left( \tau \right) = 0.$ 
The third interesting regime discussed in the literature \cite{Kovchegov:2007pq},
is the limiting case $s=0$ with $p_{\perp} \left( \tau 
\right) = -p_{\parallel} \left(\tau \right)$ which  will be adressed in 
section 5.} of $s$ this can be basically done. The crucial 
observation of Ref.~\cite{US1} is the existence of a scaling variable $v = z \,
\tau^{ - s/4}$, which captures the large 
proper-time behavior on the gravity side (see section (\ref{naiveET}) for more
details). Taking the $\tau\! \to\! \infty$ limit while 
keeping the scaling variable fixed reduces the Einstein's equations
\eqref{einsteinequations} for the functions $a\left(\tau, z \right)$, 
$b\left(\tau, z \right)$ and $c\left(\tau, z \right)$ to ordinary nonlinear differential equations for their scaling forms 
$a_{0}\left(v\right)$, $b_{0}\left( v \right)$ and $c_{0}\left( v \right)$, which can be solved exactly leading to

\begin{eqnarray}
\label{scalingSOL}
a_{0} \left( v \right) &=& A\left( v \right) - 2 m\left( v \right) \mathrm{,}\nonumber \\
b_{0} \left( v \right) &=& A\left( v \right) +\left(2 s - 2\right) m\left( v \right)\mathrm{,}\nonumber \\
c_{0} \left( v \right) &=& A\left( v \right) + \left(2 - s\right)m\left( v \right) \mathrm{,}
\end{eqnarray}
where
\begin{eqnarray}
A\left( v \right) &=& \frac{1}{2} \left\{ \log{\left( 1 + \Delta\left(s\right) \, v^{4} \right)} +  \log{\left( 1 - 
\Delta\left(s\right) \, v^{4} \right)}\right\}\mathrm{,}\nonumber\\
m\left( v \right) &=& \frac{1}{4 \Delta \left( s \right)} \left\{ \log{\left( 1 + \Delta\left(s\right) \, v^{4} \right)} -  \log{\left( 
1 - \Delta\left(s\right) \, v^{4} \right)}\right\}
\end{eqnarray}
with
\begin{equation}
\Delta\left( s \right) = \sqrt{\frac{3 s^{2} - 8 s + 8}{24}}\ \mathrm{.}
\end{equation}
It turns out however, that the geometry is generically singular for all the
values of $s$, apart from $s = 4/3$. This can be checked by 
evaluating the square of the Riemann tensor (\emph{i.e.} Kretschmann scalar) in the scaling limit and requiring its regularity (the authors of 
Ref.\cite{Kovchegov:2007pq} made also the observation that $s = 4/3$ is the only value for which the functions (\ref{scalingSOL}) 
are single-valued).

For $s = 4/3$ the energy-momentum tensor takes asymptotically the form

\begin{equation}
T_{\mu \nu} = \mathrm{diag} \left\{ \frac{\#}{\tau^{4/3}} + \ldots, \, \tau^{2} \, \frac{1}{3} \, \frac{\#}{\tau^{4/3}} + 
\ldots,\frac{1}{3} \, \frac{\#}{\tau^{4/3}} + \ldots,\, \frac{1}{3} \, \frac{\#}{\tau^{4/3}} + \ldots \right\} \mathrm{,}
\label{expansion}\end{equation}
where $\#$ denotes some numerical constant. From this expression it is clear that $p_{\parallel}\left(\tau\right) = p_{\perp}\left(\tau\right)$ 
(asymptotically), the system is in local equilibrium and is described by the perfect fluid hydrodynamics. The gravity dual in this 
scaling limit is given then by

\begin{equation}
\label{perfectFLUID}
\mathrm{d} s_{\tau \to \infty}^{2}\!\! = - \frac{\left(1\! -\! z^{4} \cdot \tau^{-4/3} \right)^{2}}{z^{2}\left(1 \!+\! z^{4} \cdot 
\tau^{-4/3}\right)} \mathrm{d} \tau^{2} \! +\! \frac{1}{z^{2}} \tau^{2} \left( 1 \!+\! z^{4} \cdot \tau^{-4/3} \right) \mathrm{d} 
y^{2}\!+\!  \frac{1}{z^{2}} \left( 1 \!+\! z^{4} \cdot \tau^{-4/3} \right) \mathrm{d} \mathrm{\textbf{x}}_{\perp}^{2} \!+ \! 
\frac{1}{z^{2}} \mathrm{d} z^{2}
\end{equation}
and looks like a boosted and dilated black brane 
\begin{equation}
\mathrm{d} s^{2} = - \frac{\left( 1 - z^{4} \lambda^{4} \right)^{2}}{z^{2} \left( 1+z^{4} \lambda^{4} \right)} u_{\mu} \, u_{\nu} \, 
\mathrm{d}x^{\mu} \, \mathrm{d} x^{\nu} + \frac{1}{z^{2}} \left(1 + z^{4} \lambda^{4}\right) \left( \eta_{\mu \nu} + u_{\mu} u_{\nu} 
\right) \mathrm{d} x^{\mu} \mathrm{d} x^{\nu}+ \frac{1}{z^{2}} \mathrm{d} z^{2}
\end{equation}
with boost parameter $u^{\mu} = 1 \cdot \left[\partial_{\tau}\right]^{\mu}$ and dilatation $\lambda \sim \tau^{-1/3}$. Recently this connection has been 
exploited and formulated as a \emph{fluid/gravity duality} \cite{MINWALLA}. Note that \eqref{perfectFLUID} contains a  $(\tau \tau)$ 
coefficient
which goes to zero at $z\sim \tau^{1/3},$ giving a hint for 
a black brane  horizon \cite{Figueras:2009iu}.

It is also clear that the metric (\ref{perfectFLUID}) is not an exact solution of
Einstein's equations -- there are subleading
effects not captured by the scaling variable limit. Those with
power-like scaling correspond to dissipative corrections in hydrodynamics.

\subsection{Large proper-time expansion and hydrodynamics}

When evaluating $\mathcal{R}^{ 2}\equiv\mathcal{R}_{A B C D} \mathcal{R}^{A B C
D}$ in the scaling limit ($\tau \rightarrow 
\infty$  keeping $v = z \cdot \tau^{-1/3}$ fixed) on the asymptotic solution,
the following pattern is encountered

\begin{equation}
\label{RiemSqLEADING}
\mathcal{R}^{ 2} = (\mathrm{nonsingular})\mathcal{R}^{ 2}_{\left(0\right)}
\left( v \right) + \frac{1}{\tau^{4/3}} \cdot 
(\mathrm{singular})\mathcal{R}^{ 2}_{\left(4/3\right)} \left( v \right) + \ldots
\end{equation}
where $(\mathrm{singular})\mathcal{R}^{ 2}_{\left(4/3\right)}\left( v \right)$
has a fourth-order pole at $v = 3^{1/4}$ and the 
dots ``$\ldots$'' denote terms suppressed by inverse powers of $\tau$ higher
than $4/3$. Any background with singularities not covered 
by the event horizon would correspond to unphysical configurations on the gauge
theory side (see e.g. \cite{Janik:2008tc, 
Buchel:2000ch}). Thus the condition of nonsingularity is the crucial
requirement, which singles out the correct behavior of the 
geometry and thus of the energy-momentum tensor of the boundary plasma. It is
obvious that the singularity at $\tau^{-4/3}$ in 
(\ref{RiemSqLEADING}) can be cancelled only if there is an additional
contribution to the metric warp factors which arises from a correction to the
perfect fluid asymptotics
\eq
\eps(\tau)=\f{1}{\tau^{\f{4}{3}}} \left(1 +\f{\#}{\tau^r}+\ldots \right)
\eqx
Assuming now the expansion
\begin{equation}
a \left( \tau, z \right) = a_{0} \left( z \cdot \tau^{-1/3} \right) +
\frac{1}{\tau^{r}} \, a_{1} \left( z \cdot \tau^{-1/3} \right)
+ \frac{1}{\tau^{4/3}} \, a_{2} \left( z \cdot \tau^{-1/3} \right) + \ldots
\end{equation}
with analogous equations for the other warp factors $b$ and $c$ and  solving
Einstein's equations at orders $\tau^{-r}$ and $\tau^{-4/3}$ 
for $r \neq 2/3$ one ends up with a singular Kretschmann scalar $\mathcal{R}^{
2}$ of the following form
\begin{eqnarray}
\mathcal{R}^{ 2}&=& (\mathrm{nonsingular})\mathcal{R}^{
2}_{\left(0\right)}\left( v \right) + \frac{1}{\tau^{r}} \cdot 
(\mathrm{nonsingular})\mathcal{R}^{ 2}_{\left(r\right)} \left( v \right) + \\
\nonumber
&& + \frac{1}{\tau^{4/3}} \cdot (\mathrm{singular})\mathcal{R}^{
2}_{\left(4/3\right)} \left(v\right)+ \frac{1}{\tau^{2 r}} 
\cdot (\mathrm{singular})\mathcal{R}^{ 2}_{\left(2 r\right)}\left( v \right)
\end{eqnarray}
However for $r = 2/3$ the singular contribution at $\tau^{-2 r}$ cancels the
singularities at $\tau^{-4/3}$ (see \cite{Heller:2007qt} 
for details) and gives a regular Kretschmann scalar up to the terms of order
$\tau^{-4/3}$ in the scaling limit. 
Physically, on the gauge theory side, this exactly corresponds to the first
viscous corrections to the perfect fluid expansion.
This argument repeated 
at any higher order fixes the large proper-time expansion of the bulk metric to
take the form

\begin{eqnarray}
a \left( \tau, z \right) &=& a_{0} \left( z \, \tau^{-1/3} \right) +
\frac{1}{\tau^{2/3}} \, a_{1} \left( z \, \tau^{-1/3} \right) + 
\frac{1}{\tau^{4/3}} \, a_{2} \left( z \, \tau^{-1/3} \right) + \ldots \nonumber
\\
b \left( \tau, z \right) &=& b_{0} \left( z \, \tau^{-1/3} \right) +
\frac{1}{\tau^{2/3}} \, b_{1} \left( z \, \tau^{-1/3} \right) + 
\frac{1}{\tau^{4/3}} \, b_{2} \left( z \, \tau^{-1/3} \right) + \ldots \nonumber
\\
c \left( \tau, z \right) &=& c_{0} \left( z \, \tau^{-1/3} \right) +
\frac{1}{\tau^{2/3}} \, c_{1} \left( z \, \tau^{-1/3} \right) + 
\frac{1}{\tau^{4/3}} \, c_{2} \left( z \, \tau^{-1/3} \right) + \ldots 
\end{eqnarray}
The functions $a_{i}\left( v\right)$, $b_{i}\left( v\right)$ and $c_{i}\left(
v\right)$ can be obtained order by order solving Einstein's 
equations in the scaling variable. At each order $i > 0$ there is a single
integration constant fixed neither by the equations of motion 
nor the asymptotic behavior of the metric, which is a combination of various
transport coefficients of the plasma. Each 
constant\footnote{The constant $\eta_0$ in first order,
$\frac{\lambda^{\left(0\right)}_{1}}{3 \sqrt{3} \, \pi^{2}} - \frac{\eta_{0}
  \tau^{\left(0\right)}_{\Pi}}{3 \sqrt{3} \, \pi^{2}}$ in the second order,
\ldots} in the expansion of the energy density or 
equivalently the temperature ($\epsilon \sim T^{4}$)
\begin{equation}
\label{energyDEN}
T\left( \tau \right) = \frac{\Lambda}{\tau^{1/3}} \left\{1 -
\frac{1}{\Lambda \, \tau^{2/3}} \cdot \frac{\eta_{0}}{\sqrt{2} \, 3^{1/4}
  \pi}+ \frac{1}{\Lambda^{2} \, \tau^{4/3}} \left(
\frac{\lambda^{\left(0\right)}_{1}}{3 \sqrt{3} \, \pi^{2}} - \frac{\eta_{0}
  \tau^{\left(0\right)}_{\Pi}}{3 \sqrt{3} \, \pi^{2}} \right)+ \ldots
\right\}
\end{equation}
is fixed uniquely by requiring the nonsingularity of the geometry at one order
higher. In the above equation $\Lambda$ denotes the 
overall energy scale defined by $T = \Lambda \cdot \tau^{-1/3} + \ldots$,
$\eta_{0}$ comes from the shear viscosity, while
$\lambda_{1}^{(0)}$ and $\tau_{\Pi}^{(0)}$ are related to second order
transport coefficients $\lambda_1$ and relaxation time $\tau_{\Pi}$
(see \cite{Baier:2007ix}). The expansion in 
$\tau^{-2/3}$ can be easily understood using hydrodynamics (see e.g.
\cite{MINWALLA} for a detailed discussion), which is an effective 
description of systems at local equilibrium described by four degrees of
freedom: the four-velocity $u^{\mu}$ ($u_{\mu} \, u^{\mu} = 
-1$) and the temperature $T$. Those quantities vary over space-time, however the
scale set by their variations is much bigger than the 
microscopic scale given by the temperature. Thus the small parameter in this
case is $T^{-1} \nabla_{\mu} u^{\nu}$. In a 
boost-invariant setup the asymptotic scaling of the temperature obtained from
the nonsingularity is $\tau^{-1/3}$, the covariant 
derivative of the velocity\footnote{In a boost-invariant setup the fluid
velocity takes the form $u^{\mu} = 1 \cdot 
\left[\partial_{\tau}\right]^{\mu}$, but the metric in $\tau$ - $y$ coordinates
has nonvanishing Christoffel symbols leading to 
$\nabla u = \tau^{-1}$.} scales as $\tau^{-1}$ which leads to the expansion
parameter $\tau^{-2/3}$ in exact agreement with 
the one obtained on the gravity side. For phenomenological reasons (from
causality, see the discussion in $e.g.$ \cite{Baier:2007ix}) 
it is interesting to focus on an energy density expanded up to the second order
in $\tau^{-2/3}$ expansion, so that the geometry has to 
be solved up to the third order to fix all the constants of interest.

In \cite{Heller:2007qt} a surprising feature showed up: the pole contributions
to the  Kretschmann scalar  at the third order 
cancelled, leaving a logarithmic singularity of the form $\log{\left(3^{1/4} - v
\right)}$.  This  singularity was present also in 
higher order curvature invariants  in Fefferman-Graham variables, such as 
$\mathcal{R}_{A B C D} \mathcal{R}^{C D E F} \mathcal{R}_{E
F}^{\,\,\,\,\,\,\,\,\, A B}$, which led to the conjecture that the gravity 
dual to boost-invariant flow cannot be realized within the supergravity
approximation \cite{Benincasa:2007tp,Buchel:2008xr}. The 
resolution of this puzzle involves resummation of the series

\begin{equation}
\mathcal{R}^{ 2} = \mathcal{R}^{ 2}_{0} \left( v \right) + \frac{1}{\tau^{2/3}}
\mathcal{R}^{ 2}_{1} \left( v 
\right) + \frac{1}{\tau^{4/3}} \mathcal{R}^{ 2}_{2} \left( v \right) +
\frac{1}{\tau^{2}} \mathcal{R}^{ 2}_{3} \left(v 
\right) + \ldots
\end{equation} 
which turns out to be equivalent to the change of variables from
Fefferman-Graham to Eddington-Finkelstein metric \cite{Heller:2008mb}. 
Subsequently it was shown, that the geometry dual to the boost-invariant flow
in the large $\tau$ limit is well-defined at any order of 
$\tau^{-2/3}$ expansion \cite{Kinoshita:2008dq} and the usual black brane
singularity is covered by an event horizon 
\cite{Figueras:2009iu, BoothHellerSpalinski}. Thus the universal behavior of
boost-invariant plasma at large proper-time (note that the 
energy density contains only one arbitrary constant, which sets the overall
scale) possesses a regular bulk description, which could 
be constructed analytically in a perturbative manner. Moreover, as expected, it
does not matter which coordinate frame in the bulk is 
used, so that in the rest of the paper all the calculations will be performed
in Fefferman-Graham coordinates.

\section{Early time dynamics\label{sectionET}}

\subsection{No scaling at early time\label{naiveET}}

The second regime where analytic methods can be applied is the domain of early
times. At first sight it seems that the early-time problem could be addressed by
the 
same method which was so useful for late time, namely solving the 
boost-invariant Einstein's equations \eqref{einsteinequations} 
without specifying {\it a priori} initial conditions. However after careful
consideration a crucial difference turns out to occur: no scaling solution 
appears which would consist of a 
hierarchy of terms similar to \eqref{expansion} and independent of the initial
conditions. In fact, on a more physical 
ground, the initial conditions should play a crucial role at early times.  This
would indeed agree with the physical intuition which suggests that this regime
should not be universal, but rather depend on initial conditions. 
However, as discussed in the following, the solutions are submitted
nevertheless to nontrivial constraints
if they are considered in the strong coupling limit.

As will be shown in the rest of the paper the search for a scaling variable (as in 
Ref. \cite{Kovchegov:2007pq}) is  invalidated due to a subtlety. In order to see
this one needs to recall the way how to identify the existence of a scaling
variable. Consider the leading proper time dependence of the energy density in
the 
regime of interest ($\tau \to \infty$ or $\tau \to 0$)
\eq
\eps(\tau) \sim \f{1}{\tau^s}
\eqx  
and then solve the Einstein's equations  \eqref{einsteinequations}
\emph{exactly} as a power series around the boundary
\eq
g_{\mu \nu} = \eta_{\mu \nu} + g_{\mu \nu}^{\left( 4 \right)}(\tau) \, z^{4} + g_{\mu \nu}^{\left( 6 \right)}(\tau) \, z^{6} + g_{\mu 
\nu}^{\left( 8 \right)}(\tau) \, z^{8} + \ldots
\eqx
The coefficients of this power series are  explicit $s$-dependent functions of
proper time $\tau$ and typically contain a couple of 
terms. If the part of $g_{\mu \nu}^{\left( n \right)}(\tau)$ which dominates at
$\tau \to \infty$ is selected, the above power series 
becomes a power series of a scaling variable $v=z/\tau^{\f{s}{4}}$. If on the
other hand the parts that dominate for $\tau \to 0$ are 
considered, it leads \cite{Kovchegov:2007pq} to an expression of the form
\eq
\f{z^4}{\tau^s}\cdot f\left( w\equiv\f{z}{\tau} \right)
\eqx 
for the metric coefficients. One consequently finds a unique solution in $w$
for each $s$ with a complex branch cut singularity for 
$s>0$. This leads to the only possible value $s=0$ right at the margin of the allowed range, i.e.
\eq
\eps(\tau) \sim const  \quad\quad\quad \text{for} \quad \tau \to 0
\eqx
out of the range of generic $s$ \cite{Kovchegov:2007pq}.

However the situation when $s=0$ is special.  
Indeed, going back to the derivation of the early time scaling variable 
$w=z/\tau$, it turns out that the terms in $g_{\mu \nu}^{\left( n
\right)}(\tau)$ which lead to it are all multiplied by a factor of 
$s$. Hence for $s=0$, these terms vanish and a completely different hierarchy
of terms appear.

Indeed, if one performs for instance the power series expansion of $a(\tau,z)$
starting from
$\epsilon(\tau)=1/\tau^s$ then the answer for the first three orders
is
\eqn
-&{z}^{4}&\ {\tau}^{-s} + {z}^{6}\ \left\{ \f{1}{6}\,{\tau}^{-s-2}s-\f{1}{12}\,
{\tau} ^{-s-2}{s}^{2} \right\}+\nn
+&{z}^{8}&\ \left\{
-\f{1}{16}\,{\tau}^{-2\,s}{s}^{2}-\f{1}{6}\,{\tau}^{-2\,s}+1/6\,
{\tau}^{-2\,s}s+{\frac {1}{96}}\,{\tau}^{-s-4}{s}^{2}-{\frac {1}{384}}\,{\tau}
^{-s-4}{s}^{4} \right\} + \ldots \ .
\label{subtelty}
\eqnx
Consider now the term proportional to $z^8$ in \eqref{subtelty}. There are two structures
$z^8/ \tau^{(s+4)}$ and $z^8/\tau^{2s}$. The first of these leads to
the early-time scaling variable proposed in \cite{Kovchegov:2007pq} since it
dominates for nonzero
$s$. However its coefficient is proportional to $s$ so for $s=0$ it is
absent and the only contribution comes from the second term which is not
contained in the scaling-variable analysis. The same applies obviously already to the
term in $z^6/ \tau^{(s+2)}$ as well as to all subsequent orders.

Analysing the power series solutions in more detail, one finds that for a
generic early time expansion of the energy density\footnote{The 
restriction to even powers of $\tau$ is discussed in one of the following sections.}
\eq
\label{e.powereps}
\eps(\tau)=\sum_{n=0}^\infty \eps_{2n} \tau^{2n},
\eqx
the coefficients $a_n$ of the power series expansion in $z$ of the metric
coefficients at $\tau=0$
\eq
\label{e.powera}
a(\tau=0,z)=\sum_{n=0}^\infty a_n z^{4+2n}
\eqx
depends on all coefficients $\eps_{2n}$ in \eqref{e.powereps}. Reformulating
this observation, this means that each of the possible 
initial conditions (\ref{e.powera}) leads to a distinct proper-time evolution
(\ref{e.powereps}), where the coefficients of the two 
power series are linked through the Einstein's equations. The following mapping
is therefore obtained
\eq
a(\tau=0,z)=\sum_{n=0}^\infty a_n z^{4+2n}  
\quad\quad \Longrightarrow \quad\quad
\eps(\tau)=\sum_{n=0}^\infty \eps_{2n} \tau^{2n} \mathrm{.}
\eqx
On the other hand, certain nontrivial constraints limit the
range of solutions. Among these, the nonsingularity condition 
on the metric-invariant properties of the geometry plays the crucial role.

The derivation of this result together with the constraints including the
nonsingularity argument restricting the early time expansion 
in terms of  even powers of the proper time of the energy density, is the main
result of this paper and is presented in section 
(\ref{enDENfromMETRIC}).

The objective now is to analyze the qualitative properties of the evolution
of the energy density starting from the initial 
conditions (\ref{e.powera}). But before this can be done, the space of allowed
initial conditions has to be investigated in more 
detail, as these have to satisfy a nonlinear constraint equation which is a
part of the Einstein's equations \eqref{einsteinequations}.

\subsection{The Fefferman-Graham metric is  singular at all
times\label{INITcond}}

This section focuses on the restrictions on the initial conditions coming from
Einstein's equations, which can be written in an
equivalent simplified form as
\eq
\label{EINSTEINeqns}
R_{AB}+4 G_{AB}=0
\eqx
or explicitly in Fefferman-Graham coordinates as

\begin{eqnarray}
\label{einsteinequations}
&(\tau \tau)&: {\ddot b} \!+\! 2 {\ddot c}\!-\!\f{\dot a}2(\dot b\! +\!2 \dot
c)\! +\!
\f 12 ({\dot b}^2 \!+\! 2{\dot c}^2 )\!-\!\f 1\tau({\dot a}\!-\!2{\dot b}) =
e^a\left\{a''\!-\!\f{3a'}z\!+\!\left(\f {a'}2\!-\!\f
1z\right)(a'\!+\!b'\!+\!2c')\right\} \mathrm{,}\nonumber \\
&(y y)&:{\ddot b}\!-\!{\dot a}{\dot b} \!+\!\f1\tau (\dot b\! -\!2 \dot a)\! +\!
\f 12 ({\dot a} \!+\!{\dot b}\!+\! 2{\dot c}) \left({\dot b}\!+\!\f 2\tau\right)
= e^a\left\{b''\!-\!\f{3b'}z\!+\!\left(\f {b'}2\!-\!\f
1z\right)(a'\!+\!b'\!+\!2c')\right\} \mathrm{,}\nonumber \\
&(\perp \perp)&: {\ddot c}-{\dot a}{\dot c}  +\f{\dot c} 2\left({\dot a}+{\dot
b}+2{\dot c}+\f 2\tau\right) = e^a\left\{c''-\f{3c'}z+\left(\f {c'}2-\f
1z\right)(a'+b'+2c')\right\} \mathrm{,}\nonumber \\
&(\tau z)&: 2 {\dot b}' + 4 {\dot c}'+b'\left({\dot b}\!+\!\f 2\tau\right)+2\dot
c c'- a'\left({\dot b}+2 {\dot c}+ \f 2\tau\right)=0  \mathrm{,}\nonumber \\
&(z z)&:  a'' + b'' +2 c'' -\f 1z(a'+b'+2c') +\f12(a'^2+b'^2 +2c'^2) =0
\mathrm{.}
\end{eqnarray}
In the above expressions the first parenthesis is for the corresponding metric
component, the dot
for the proper-time $\tau$-derivative and the prime for the $z$-derivative.

Expanding the $({z z})$ component of Einstein's equations as a power series in $z$, and keeping only the leading terms for $z\rightarrow 0$, one obtains the tracelessness condition $-\epsilon\left( \tau \right)+p_{\parallel} \left( \tau \right)+2 p_{\perp} \left( \tau \right)=0$ of $\cor{ T_{\mu \nu} }$. The $({z z})$ equation thus encodes holographically the scale invariance of the boundary gauge theory. Moreover, that Einstein's equation does not features any $\tau$-derivative. Hence, at each proper time, if the $z$-dependence of two of the functions $a$, $b$ and $c$ are given, the third one is completely determined by the $({z z})$ equation and the requirement of a $z^4$ behavior near the boundary.

However, for completeness, let us discuss also some of the other Einstein's equations. Expanding the $({z \tau})$ component of Einstein's equations as a power series in $z$, one finds that the leading terms near the boundary give the relation ${\dot p_{\parallel}}+2{\dot p_{\perp}}+(\epsilon+p_{\parallel})/\tau=0$ which, assuming $\cor{ T_{\mu}^{\mu}}=0$, is equivalent to the energy conservation equation in the CFT. Hence, the $({z\tau})$ and $({z z})$ components of Einstein's equations together encodes holographically both energy conservation and conformal invariance in the CFT.

One should remark that we have five equations \eqref{einsteinequations} for only three unknown functions. There is indeed some redundancy in Einstein's equations \eqref{einsteinequations}. It is \emph{e.g.} consistent to replace the first three equations $({\tau \tau})$, $(y y)$ and $(\perp \perp)$ by the combination $-({\tau \tau})+(y y)+2(\perp \perp)-e^a (z z)$, which writes
\eq
\label{combinedEinsteinEq}
{\dot c}\left({\dot c}+2{\dot b}+\frac{4}{\tau} \right)=e^a \left\{2b''+4 c''+b'^2+3 c'^2+2 b'c'- \frac{6}{z} b'- \frac{12}{z} c' \right\} \mathrm{.}
\eqx
Hence, the system of Einstein's equations is only of order one in $\tau$. The required initial conditions should thus be two of the three functions $a_{0}(z)$, $b_{0}(z)$ and $c_{0}(z)$, where $a_{0} = a_{0} \left(z\right) = a \left( \tau = 0,z\right)$ and analogous expressions for the others, because we have already at $\tau=0$ the constraint equation 
\eq
\label{constraint1}
a_0''+b_0''+2 c_0'+\f{1}{2} (a'_0)^2 +\f{1}{2} (b'_0)^2+ (c'_0)^2- \f{1}{z} \left(a'_0 +b'_0+2 c'_0\right)=0 \mathrm{,}
\eqx
We shall see in the next section that requiring a regular behavior of the solutions in the limit $\tau\rightarrow 0$ further reduces the freedom for the initial condition to only one arbitrary function of $z$.

 The nonlinear character of equation \eqref{constraint1}
turns out to play a crucial role in the whole analysis. Naively, one would
expect to be able to consider a small fluctuation over empty $AdS_5$ in the
linearized approximation neglecting the quadratic terms in the above
constraint. However it turns out, as will be shown below, that even if starting
from $e.g.$ an infinitesimal $a$, the nonlinear equation
for $c$ will always generate a singularity for some large but finite $z$! Thus
linearized fluctuations cannot be used as initial
conditions in the boost-invariant kinematics. Moreover, a singularity,
presumably related to an horizon, has to be present from the
outset already in the initial conditions. This point will be addressed later in the paper.

To simplify \eqref{constraint1} it is useful to introduce the following notation
\eqn
v(z^2) &=& \frac{1}{4z} a_0'(z)  \mathrm{,} \nonumber\\
\varsigma(z^2) &=& \frac{1}{4z} b_0'(z)  \mathrm{,} \nonumber\\
w(z^2) &=& \frac{1}{4z} c_0'(z) \mathrm{.}
\eqnx
If the constraint equation has a regular solution, the derivatives $a_0',$
$b_0'$ and $c_0'$ are bounded in the bulk and $v,\varsigma$ and $w$ vanish when $z \to
\infty$. The constraint equation simplifies when written in terms of $v$, $\varsigma$
and $w$
\eq
\label{e.constsimp}
v'+\varsigma'+2 w'+v^2+\varsigma^2+2 w^2=0 \mathrm{,}
\eqx
where the prime now means the $z^2$-derivative.
Integrating the above equation assuming the regularity conditions from $z=0$ to
$\infty$ gives
\eq
\int_0^\infty (v'+\varsigma'+2 w') dz^2+\int_0^\infty (v^2+\varsigma^2+2 w^2)dz^2=0 \mathrm{.}
\eqx
The first integral vanishes due to the imposed boundary conditions leading to
\eq
\int_0^\infty (v^2+ \varsigma^2+2 w^2) dz^2=0 \mathrm{.}
\eqx
This equality however is satisfied only for $v=\varsigma=w=0$, thus the only regular
solution is trivial -- the vacuum $AdS_{5}$. Therefore, {\it at any time} the
metrics of interest must have a singularity at some value of $z.$ In particular
this will be the case even at $\tau=0,$ with the meaning that any nontrivial
initial condition consistent with the Einstein's equations will lead to a metric
singularity at some value of $z$. The nonsingularity constraint on
the geometry requires that all the singularities apart from the one sitting at
$z = \infty$ will only be of coordinate nature. This provides a strong
selection mechanism for the allowed initial conditions.

\subsection{Constraints at early time\label{INITconstr}}

This section analyzes the constraints on initial conditions imposed by the Einstein's equations together with the assumption of nonsingularity of the geometry at $\tau = 0$. The $(\tau z)$ component equation (see \eqref{einsteinequations})
 takes the form
\eq
\partial_z a-\partial_z b = \tau \cdot (\ldots)\mathrm{,}
\eqx
which means that at $\tau=0$, $a(\tau,z)$ and $b(\tau,z)$ can differ only by a constant. Since both functions have to vanish at $z=0$, 
this constant vanishes as well. Therefore
\eq
\label{e.aeqb}
a(0,z)=b(0,z) \mathrm{.}
\eqx
%In the rest of the paper $a_0(z) \equiv a(0,z)$ and $c_0(z) \equiv c(0,z)$ for brevity. 
Incidentally the condition (\ref{e.aeqb}) ensures also that the geometry is
nonsingular on the light-cone $\tau=0$ for generic $z$.

In fact the constraint equation can be solved exactly leading to the space of
solutions parametrized fully by the single function. The 
trick is to introduce the linear combinations
\eqn
v_+ &=& -w-v \mathrm{,} \nonumber \\
v_- &=& w-v
\eqnx
for which the equation (\ref{e.constsimp})  becomes algebraic for $v_-$ (both $v_-$ and $v_+$ are understood as functions of $z^2$). After
trivial algebra one obtains
\eq
\label{e.vminus}
v_-= \sqrt{2v_+'-v_+^2} \mathrm{.}
\eqx
Therefore all solutions of the initial value  nonlinear constraint equations are
parametrized by an arbitrary function $v_+(z^2)$. The next step is to analyze
what further conditions must be imposed on $v_+(z^2)$. First, since the the
metric coefficient functions have to vanish as $z^4$ one gets
\eq
v_+(z^2) \sim \f{2}{3} \eps_0 z^6   \quad\quad \text{for} \quad z \sim 0 \mathrm{.}
\eqx 
Moreover, from the arguments in section 4.2 it follows that there is a
singularity at some finite $z=z_0$. The behavior at this 
singularity is constrained by demanding that this would be just a coordinate
singularity and not a curvature blow-up. Assuming a power-like blow-up of
$v_{+}\left( z^2 \right)$ at $z = z_{0}$, the regularity of the square of the
Riemann tensor leads to the conclusion, that $v_+(z^2)$ has to have a first
order pole
\eq
v_+(z^2) \sim \f{1}{z^2_0-z^2} \quad\quad \text{for} \quad z \sim z_0
\eqx
with residue 1.

The coordinate singularity in $v_{+}$ at $z=z_{0}$ translates directly into the
behavior of the metric coefficients around $z_0$. This means that the proper
time metric component has a second order zero at $z_0$ so that the metric at
$\tau=0$ looks like
\eq
ds^2 = -\f{1}{z^2} \left(1-\f{z}{z_0}\right)^2 \{d\tau^2 +\tau^2 dy^2\}+ \ldots
+ \f{1}{z^2} dz^2
\eqx
in the vicinity of $z=z_0$, reminiscent of the behavior of a horizon in
Fefferman-Graham coordinates. Note, however, that at $\tau=0$ the term in curly
braces becomes $dx^+ dx^-$ in contrast to a
`Schwarzschild' horizon where the corresponding structure is of the form
$-(1-z/z_0)^2 dt^2+\ldots$. 
Presumably the metric
singularity could be avoided by using a different set of 
coordinates (e.g. Eddington-Finkelstein ones). Then it would be also possible to
apply the framework of dynamical horizons to check and locate eventual trapped
surfaces. However performing a change of variables from Fefferman-Graham to
Eddington-Finkelstein ones is a formidable task, since contrary to late
proper-time regimes, $\tau \sim 0$ Fefferman-Graham is not mapped in the bulk
into $\tilde{\tau} \sim 0$ Eddington-Finkelstein. This means that the change of
variables cannot be performed perturbatively and is beyond the scope of
analytical methods presented here. It would be however interesting to analyze in
detail the local structure of the geometry at $z = z_{0}$ including
time-dependence. This problem is left for future work. 

It is interesting for further discussion to present explicit solutions of the
constraints. One of them arises from
the choice
\eq
v_+= a(\tan az^2 -\tanh az^2)\mathrm{.}
\label{example1}
\eqx
 The initial metric profiles may be integrated explicitly to obtain
\eqn
\label{initialPROFILEex}
a_0(z) &=& 2\log \cos a z^2 \mathrm{,} \nonumber \\
c_0(z) &=& 2\log \cosh a z^2 \mathrm{.}
\eqnx 
The above solution possesses a (coordinate -- not leading to any curvature
singularities) singularity at
\eq
z_0=\sqrt{\f{\pi}{2a}} \mathrm{.}
\eqx
More generally one can  
parametrize $v_{+}$ in the following manner

\begin{equation}
v_{+} \left( z^2 \right) = \frac{2}{3} \epsilon_{0} z^2_{0} \cdot
\frac{z^{6}}{z^2_{0} - z^2} f \left( z^2 \right) \mathrm{,}
\label{example2}
\end{equation}
where $f\left( 0 \right)=1$, $f \left( z^2_{0}\right) = \frac{3}{2 \epsilon_{0}
z_{0}^{8}}$ and otherwise is a regular function of 
$z^2$ variable for $z < \infty$. One has also to ascertain that $v_-$ obtained
from (\ref{e.vminus}), does not become complex 

Once the allowed initial conditions are under control one may proceed to
find the solution of Einstein's equations starting with these initial data.

\subsection{Early time expansion of the energy density \label{enDENfromMETRIC}}

The Einstein's equations (\ref{EINSTEINeqns}) can be solved for any energy
density perturbatively in $z^{2}$. Starting with some 
arbitrary energy density $\epsilon$, the first three nontrivial terms in the
expansion of $a\left( \tau, z \right)$ warp factor take 
the form

\begin{eqnarray}
\label{warpFACTORexp}
a\left( \tau, z \right) &=& - \epsilon (\tau ) \cdot z^4 + \left\{-\frac{\epsilon '(\tau
   )}{4 \tau }-\frac{\epsilon ''(\tau )}{12}\right\} \cdot z^6 + \Big\{ \frac{1}{6} \epsilon (\tau )^2+\frac{1}{6} \tau  \epsilon '(\tau )
   \epsilon (\tau )+\frac{1}{16} \tau ^2 \epsilon '(\tau)^2\nonumber\\
   &&+\frac{\epsilon '(\tau )}{128 \tau ^3}-\frac{\epsilon ''(\tau
   )}{128 \tau ^2}-\frac{\epsilon ^{(3)}(\tau )}{64 \tau }-\frac{1}{384}
   \epsilon ^{(4)}(\tau ) \Big\} \cdot z^{8} + \cdots
      \end{eqnarray}

This power series can be extended to an arbitrary order in $z^{2}$ by solving
Einstein's equations and the only obstructions are of a 
purely computational nature. Generically terms in the expansion
(\ref{warpFACTORexp}) contain inverse powers of proper time multiplying 
the energy density and its derivatives. Assuming the energy density can be
expanded in a regular power series around $\tau = 0$, the 
singular inverse powers of proper time in (\ref{warpFACTORexp}) will be present
unless all the odd terms vanish. This requirement 
constrains the energy density in the early times domain to be a power series in
even powers of $\tau,$ namely

\begin{equation}
\label{energyDENseries}
\epsilon\left( \tau \right) = \epsilon_{0} + \epsilon_{2} \tau^{2} + \epsilon_{4} \tau^{4} + \ldots
\ .
\end{equation}

On the other hand, taking $\tau$ to be zero in equation
(\ref{warpFACTORexp}) gives the relation between the early times energy 
density and the profile of $a_{0}\left( z \right)$, which has been signaled in
subsection (\ref{naiveET}). Basically expanding the 
initial profile of the metric in the radial AdS variable near the boundary and
comparing with (\ref{warpFACTORexp}) at $\tau=0$

\begin{equation}
a\left( 0, z\right) = a_{0} \left( z \right) = - \epsilon _0 \cdot z^4 - \frac{2}{3} \epsilon_{2} \cdot z^{6} + \left( - \frac{\epsilon _4}{2}-\frac{\epsilon _0^2}{6}\right) \cdot z^8+ \ldots
\end{equation}
allows one to solve for all the $\epsilon_{2 i}$ sitting in
(\ref{energyDENseries}). This pattern continues to any order in $z$ (and 
thus $\tau$) expansion. Note also, that each parameter in the expansion of
energy density around $\tau = 0$ is an independent 
dimensionful quantity and in order to fully specify the initial conditions
infinitely many such terms are needed. This is in stark contrast to the
late time behavior, where only one dimensionful constant ($\Lambda$) appears.
Physically, this is in agreement with the fact that in the thermally
equilibrated final stages of Bjorken expansion all differences due to initial
data should have been washed away by dissipative effects and the only parameter
characterizing the flow is an overall energy scale (given e.g. as the energy
density at a certain fixed proper time $\tau_0$).

The strategy for finding the time evolution from given initial data
is to use the Einstein's equations 
to generate the expansion of the metric warp-factors for sufficiently high
order (the bigger, the better, but for technical reasons 
this has been achieved up to the order $z^{84}$ or equivalently $\tau^{80}$ in
most cases) and then for given regular (with finite 
curvature evaluated on initial data for $z < \infty$) initial profile generate
the power series for the energy density at early times. 
Thus the AdS/CFT correspondence beautifully sets the allowed initial conditions
for gauge theory dynamics using the gravitational 
description. However since the series (\ref{energyDENseries}) has generically a
finite radius of convergence, some 
resummation method is needed in order to extend it to larger proper times.

\section{Transition to the hydrodynamic regime \label{transition2HYDRO}}
\subsection{Resummation scheme for the energy density}

\vspace{10 pt}

It is desirable now to find a suitable numerical approach, relating the
solutions of the Einstein's equations \eqref{einsteinequations} to the physical
quantities, energy density and pressures, as a function of $\tau$. 

Though early time dynamics of the field theory is dictated by the initial
conditions, after certain time the system is expected to 
settle down to local equilibrium. 
%As suggested in Ref.\cite{Nastase:2005rp},
%this process could be dual to the black hole formation. 
For $\tau$ sufficiently large the plasma should exhibit the universal
hydrodynamic behavior, where the only trace of the initial conditions 
is given by the overall scale $\Lambda,$ see \eqref{energyDEN}. In order to
track the dynamics of the system with sufficient accuracy 
from $\tau = 0$ to $\tau \gg 1$, numerical methods are needed. In particular
the early-times power series for the energy density has a 
finite radius of convergence and a resummation is needed in order to find its
behavior for large $\tau$.

We would like to address the following questions. Firstly, whether during the
evolution from some (generic) initial data at $\tau=0$ one can observe a
passage to the asymptotic perfect fluid behaviour
\begin{equation}
\label{epsilonPF}
\epsilon \left( \tau \right) \sim \frac{1}{\tau^{4/3}} + \ldots 
\end{equation}
To be more general one may try to determine the asymptotic exponent $s$ in 
\begin{equation}
\label{energySCALINGs}
\epsilon \left( \tau \right) \sim \frac{1}{\tau^{s}} + \ldots \ ,
\end{equation}
and determine whether it is significantly different (as is expected) from the
free streaming value $s=1$.
 
As explained above, it is quite difficult to answer this question as
the energy density in the early time regime takes the form
\begin{equation}
\label{energyDENcutoff}
\epsilon \left(\tau\right) = \epsilon_{0} + \epsilon_{2} \tau^{2} + \ldots +
\epsilon_{2 N_{cut}} \tau^{2 N_{cut}} + \ldots
\end{equation}
where $N_{cut}$ is a natural number denoting the cut-off up to which the
evaluation of the energy density from the initial profile in 
the bulk has been performed. In most cases $N_{cut}=40$ and increasing this
accuracy is difficult. Moreover the series has a finite range of convergence.
In order to estimate the asymptotic exponent $s$ appearing in
(\ref{energySCALINGs}) it is convenient to express it through a logarithmic
derivative
\begin{equation}
\label{LOGder}
s = - {\mathrm{lim}}_{\tau \rightarrow \infty} \, \tau \cdot
\frac{\mathrm{d}}{\mathrm{d} \tau} \log{\epsilon\left( \tau \right)} 
\mathrm{.}
\end{equation}
and perform a Pade approximation (of order $(N_{cut},N_{cut})$) to the r.h.s.
of (\ref{LOGder}) with $\eps(\tau)$ substituted with our power series
(\ref{energyDENcutoff}):
\begin{equation}
\label{padeAPPROX}
s_{\mathrm{approx}} = \frac{s_{U}^{\left( 0 \right)} + s_{U}^{\left( 2 \right)}
\tau^{2} + \ldots +s_{U}^{\left(2N_{cut} \right)} 
\tau^{2N_{cut}}}{s_{D}^{\left( 0 \right)} + s_{D}^{\left( 2 \right)} \tau^{2} +
\ldots + s_{D}^{\left(2N_{cut}\right)} \tau^{2N_{cut}}} 
\mathrm{.}
\end{equation}
$s = 4/3$ corresponds then to the perfect fluid case and $s = 1$ to the
free-streaming scenario. Of course it is not expected that 
(\ref{padeAPPROX}) will give $s = 4/3$ exactly\footnote{Such a Pade
approximation has, by construction, a different subleading large $\tau$
behaviour from viscous hydrodynamics.}. 

\begin{figure}
\label{svalue}
    \begin{center}
\includegraphics[height=4cm]{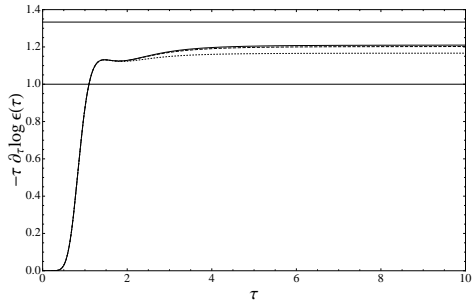}
    \caption{Approximate value of $s$ obtained from the logarithmic derivative
and Pade approximation for A) $N_{cut} = 32$ (dotted 
line), B) $N_{cut} = 40$ (dashed line) and C) $N_{cut} = 48$ (solid line) for
initial profile $v_{+}^{\left( 1\right)} \left( z^2 
\right)$. Two horizontal lines denote $s = 1$ (free streaming scenario) and $s = 4/3$ (perfect fluid case).}
    \end{center}
\end{figure}

Let us consider the three following examples of the initial profile, satisfying
the curvature  nonsingularity constraint, the first one being 
\eqref{example1}, the second its deformed variant, and the third one from the
family \eqref{example2}: 

\begin{eqnarray}
\label{profiles}
v_{+}^{\left(1\right)} \left(z^2 \right) &=& 
\tan{\left(z^2\right)} -\tanh{\left( z^2\right)} \mathrm{,}
\nonumber \\
v_{+}^{(2)} &=&  \tan{\left(z^2\right)} -\tanh{\left( z^2 + \f{z^8}{6}\right)}
\mathrm{,}\nonumber \\
v_{+}^{\left(3\right)} \left( z^2 \right) &=& \frac{2}{3} \frac{z^6}{1 -
z^2}\left(1 + \frac{1}{2} z^2\right) \mathrm{.} 
\end{eqnarray}

The obtained values of $s$ for the first profile, for which we had a power
series in $\tau$ up to order $\tau^{100}$, are $s_{\mathrm{approx}} =
1.1667, \, 1.1923, \, 1.2025, \, 1.2025$ and $1.2101$ 
respectively for $N_{cut} = 32, \, 36, \, 40, \, 44$ and $48$. The approximate
value of $s$ is closer to $s = 4/3$ than to $s = 1$ for 
largest $N_{cut}$. For the second and third profile the results are inconclusive,
since the corresponding energy densities are provided with worse 
accuracy. 

In the rest of the text, we shall keep the asymptotic perfect fluid value  $s =
4/3.$ Then, a suitable Pade resummation scheme, with 
the required asymptotia can  be given by

\begin{equation}
\label{padeAPPROXeps}
\epsilon^{3}_{\mathrm{approx}}\left( \tau \right) = \frac{\epsilon_{U}^{\left(
0 \right)} + \epsilon_{U}^{\left( 2 \right)} \tau^{2} + 
\ldots + \epsilon_{U}^{\left(N_{cut} - 2\right)} \tau^{N_{cut} -
2}}{\epsilon_{D}^{\left( 0 \right)} + \epsilon_{D}^{\left( 2 \right)} 
\tau^{2} + \ldots + \epsilon_{D}^{\left(N_{cut} - 2\right)} \tau^{N_{cut} + 2}} \mathrm{.}
\end{equation}
where both $\epsilon_{D}^{2 i}$ and $\epsilon_{U}^{2 i}$ are obtained by
expanding (\ref{padeAPPROXeps}) around $\tau = 0$ and comparing 
with (\ref{energyDENcutoff}). Such a resummation imposes the correct asymptotic
behavior, but differs, by construction, in the subleading behavior with viscous
hydrodynamics.

\begin{equation}
\epsilon_{\mathrm{approx}} \left( \tau \right) = \frac{1}{\tau^{4/3}} \left\{
\# + \frac{1}{\tau^{2}} \cdot \# \right\} \mathrm{,}
\end{equation}
(the first subleading piece scales as $\tau^{-2}$ whereas the correct scaling
is $\tau^{-2/3}$). This difference is not substantial and 
can be cured by more refined resummation schemes\footnote{An other possible
issue are roots of the denominator lying within the range 
$(0, \infty)$. If such feature is encountered, it can be interpreted as an
artifact of approximation without real physical 
significance.}.

Despite its simplicity, it turns out that the  resummation
(\ref{padeAPPROXeps}) works pretty well (the results seems to converge well 
with increasing cut-off, see  Figs.2 and 4 in extending the energy density
beyond the convergence radius of the early-time power series (which is its 
main task) and in providing a qualitative picture of dynamics.

\subsection{Qualitative features of the approach to local equilibrium}

The method (\ref{padeAPPROXeps}) can be used to study the approximate behavior
of the energy density as a function of time for large 
enough times to see local equilibration. Fig.3 shows the plots of energy
density as a function of proper time for the three  
profiles (\ref{profiles}) obtained for the highest cut-offs. The results seems
to converge  well for the first profile, see Fig.4. Energy 
densities obtained for these profiles differ at the initial stages, whereas in
the late-time regimes both seem to approach local 
equilibrium. A measure of the local equilibrium is the relative difference
between the transverse and 
perpendicular pressures defined as

\begin{equation}
\label{DELTAp}
\Delta p \left( \tau \right) = 1 - \frac{p_{\parallel} \left( \tau \right)}{p_{\perp} \left( \tau \right)}
\end{equation}

When this quantity is close to zero, it signals the isotropisation indicating
local equilibrium, while a value of order one is an 
indication in favor of the free streaming scenario.  
Fig 5 shows the plot of the relative difference of pressures \eqref{DELTAp}. It
is interesting to note the rapid fall-off of the pressure difference on a scale
$\tau={\cal O}(1).$ This fall-off appears to be stable after different numerical
checks. Interestingly enough, there is a bump which prevents the pressure  to
reach isotropy before  $\tau={\cal O}(5).$ However, the Pade approximants for
the pressure difference are less stable than for the energy density and 
the differences appear after the bump. The second profile which is a slight
deformation of the first one does not seem to exhibit this bump. 
In any case, it would be physically
interesting to check whether this phenomenon of rapid fall-off but incomplete
isotropization is or not a characteristic feature of the strong coupling
evolution. We intend to analyze this issue using numerical methods for solving
Einstein's equations in future work.

\begin{figure}
\label{fig2}
    \begin{center}

\includegraphics[width=4.85cm]{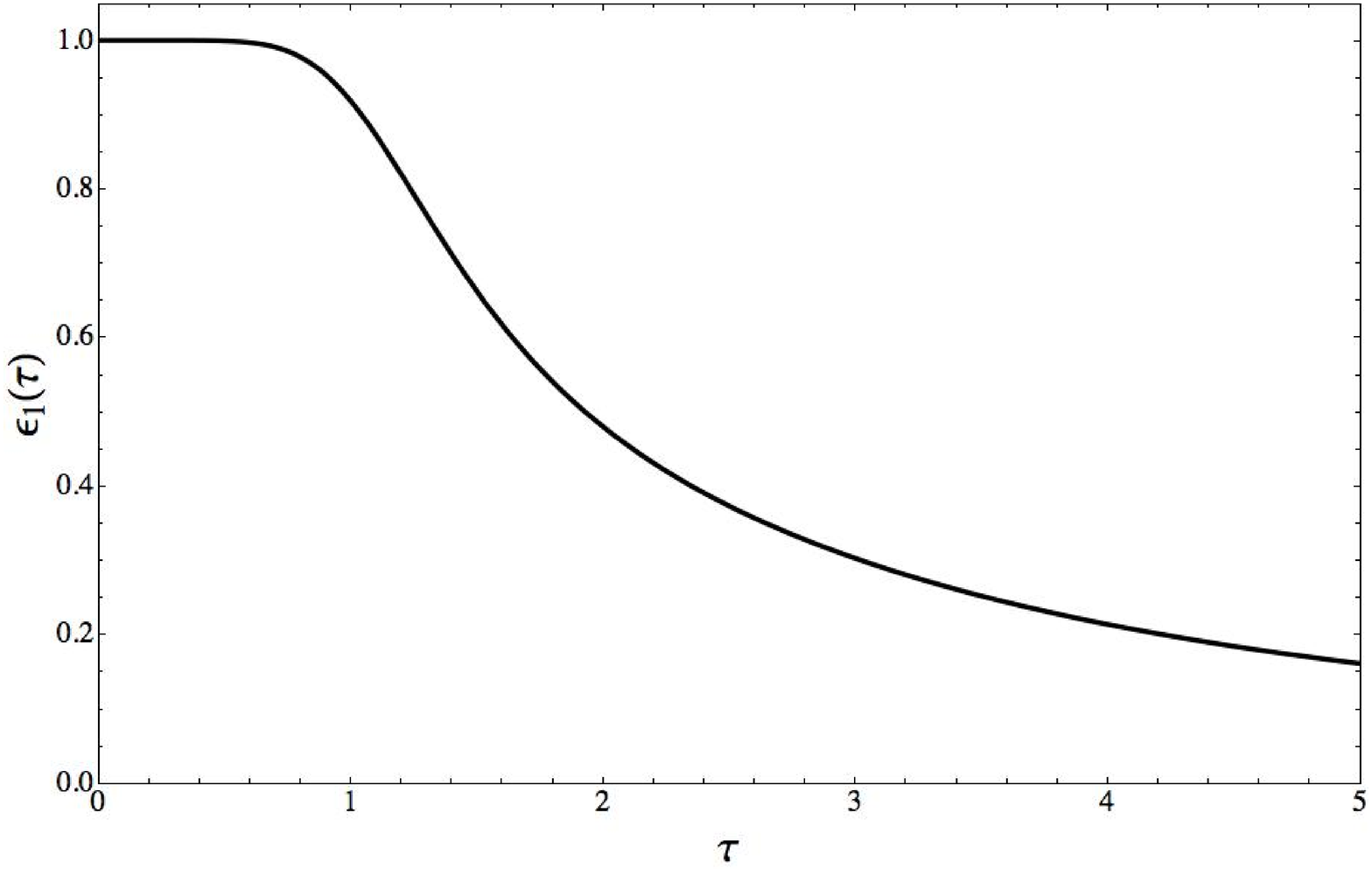}
\hspace{0.05cm}
\includegraphics[width=4.85cm]{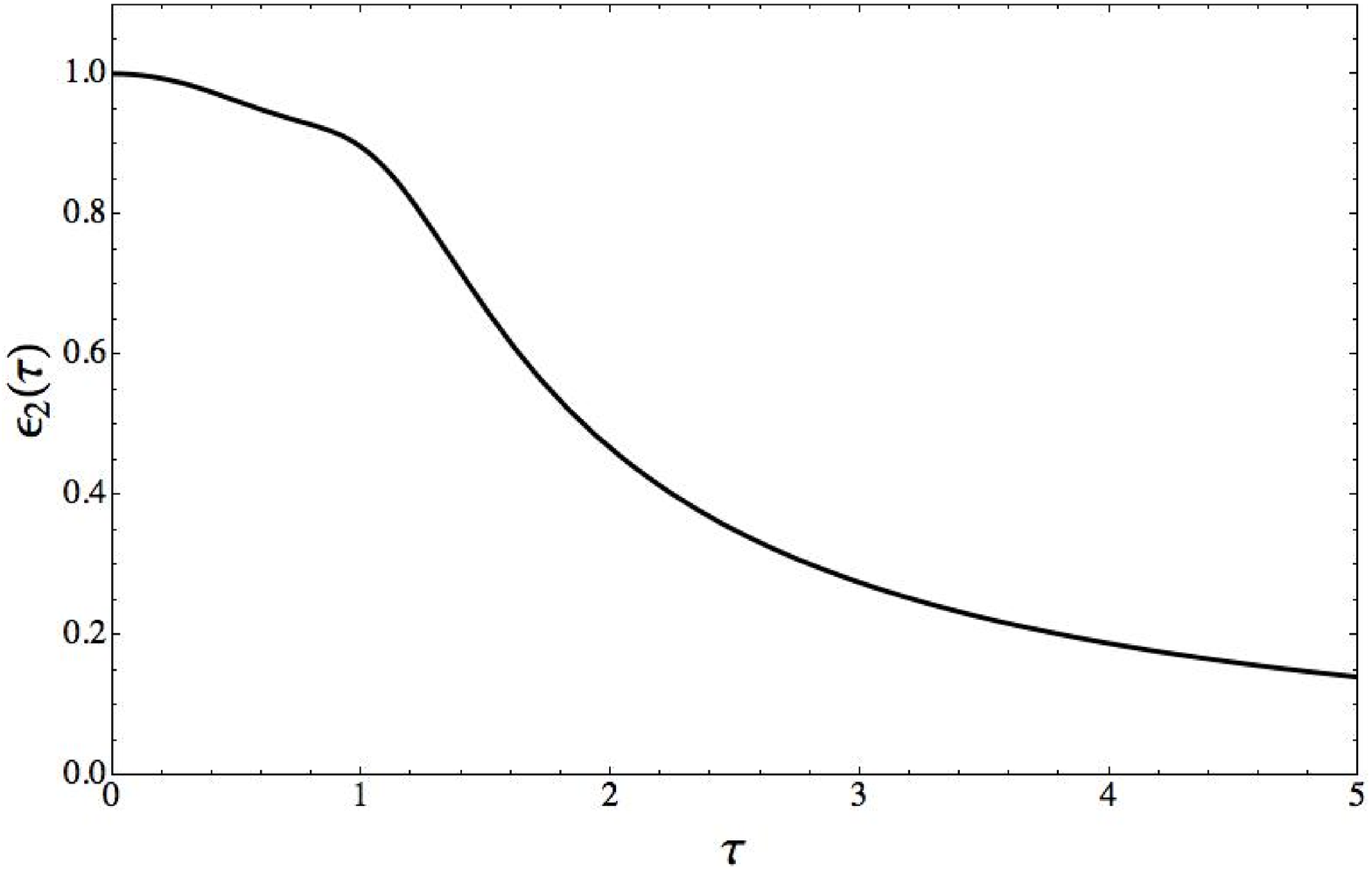}
\hspace{0.05cm}
\includegraphics[width=4.85cm]{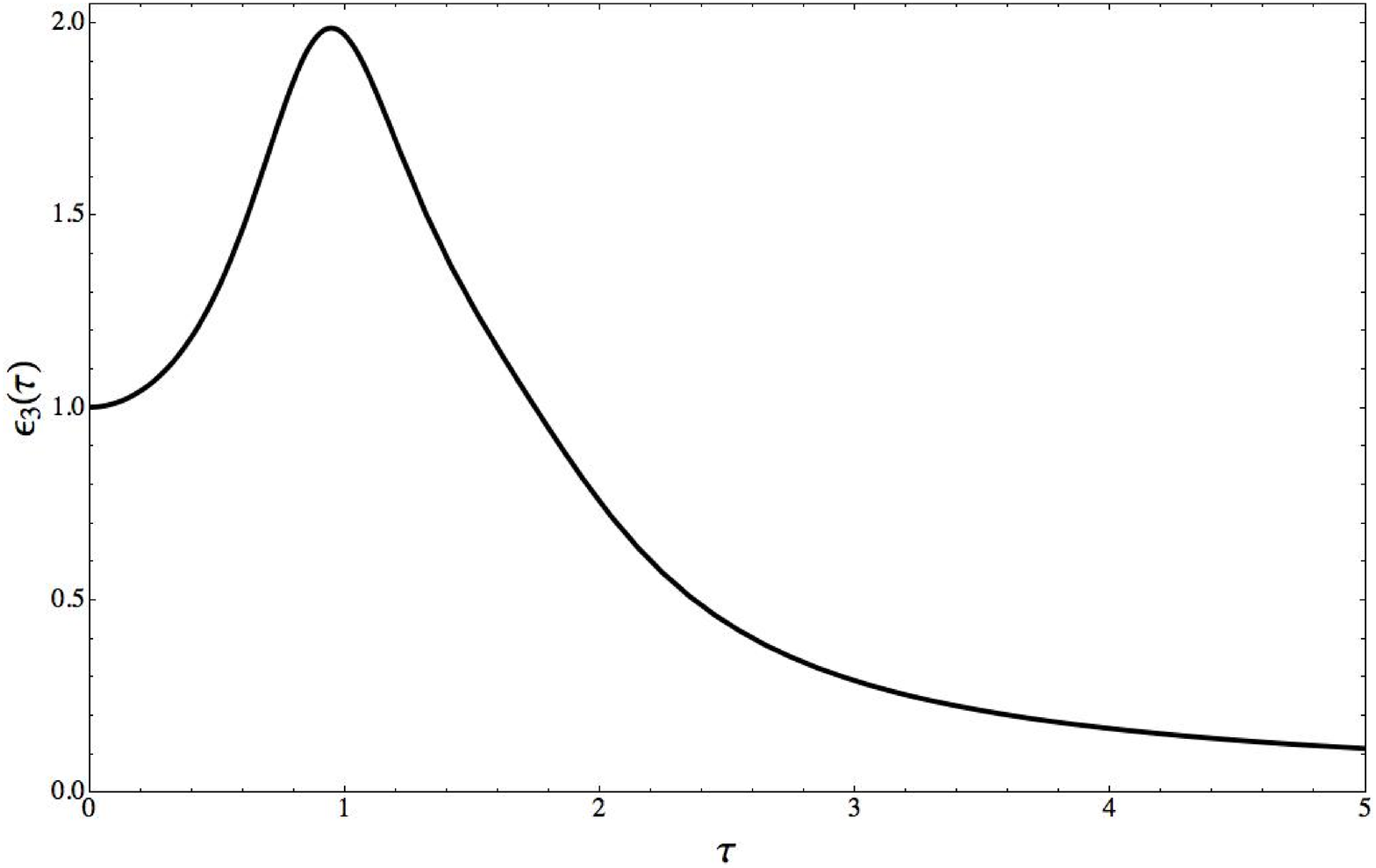}
%\vspace{-.1cm}
$$\!\!\!\!\!\!\!\! A)\ \ \ \ \ \ \ \ \ \ \ \ \ \ \ \ \ \ \ \ \ \ \  
\ \ \ \ 
\ \ \ B)\ \ \ \ \ \ \ \ \  \ \ \ \ \ \ \ \ \ \ \ \ \ \ \ \ \ \ \ \ \ \ \ C)$$
    \caption{A) Energy density $\epsilon_{1}\left( \tau \right)$ as a function
of proper-time $\tau$ obtained from Pade approximation 
for cut-off $N_{cut} = 46$ and initial profile $v_{+}^{\left( 1 \right)} \left(
z^2 \right)$ in the bulk; B) Energy density $\epsilon_{2}\left( \tau
\right)$ for the second profile. 
C) Energy density 
$\epsilon_{3}\left( \tau \right)$ as a function of proper-time $\tau$ obtained
from Pade approximation for cut-off $N_{cut} = 34$ and 
initial profile $v_{+}^{\left( 3 \right)} \left( z^2 \right)$ in the bulk.}
    \end{center}
\end{figure}

\begin{figure}
\label{fig3}
    \begin{center}
\includegraphics[height=4cm]{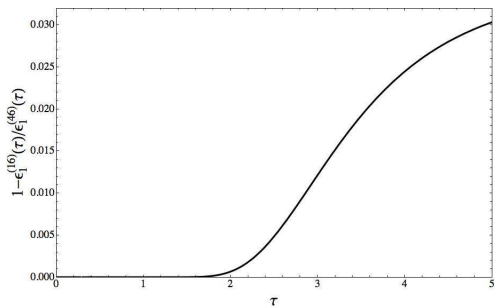}
    \caption{Relative difference between the energy densities for the first
profile for cut-offs $N_{cut} = 16$ and $N_{cut} = 46$ does not exceed 10\%.}
    \end{center}
\end{figure}

\begin{figure}
\label{fig4}
    \begin{center}
\includegraphics[height=4cm]{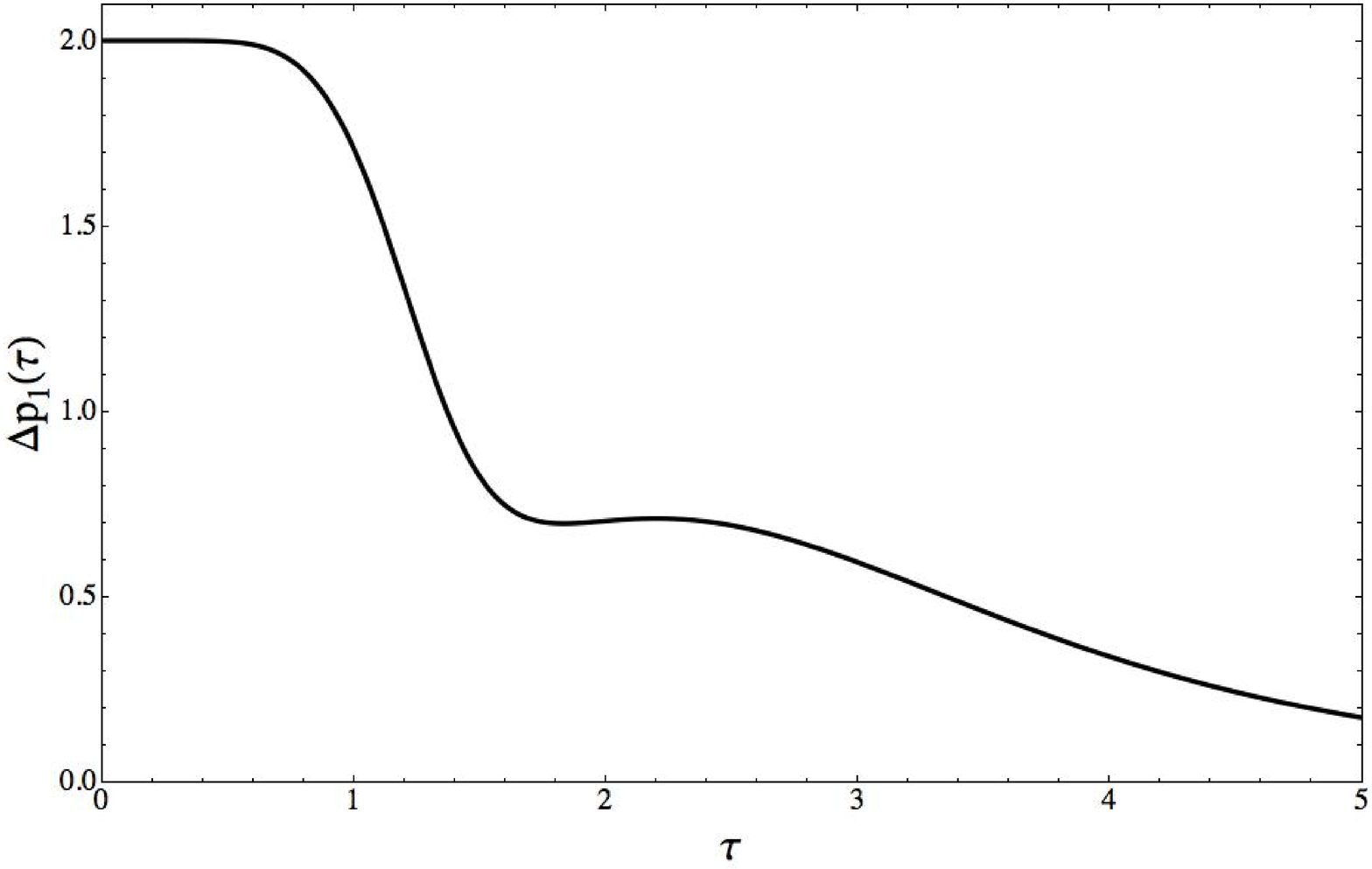}
\hspace{0.5cm}
\includegraphics[height=4cm]{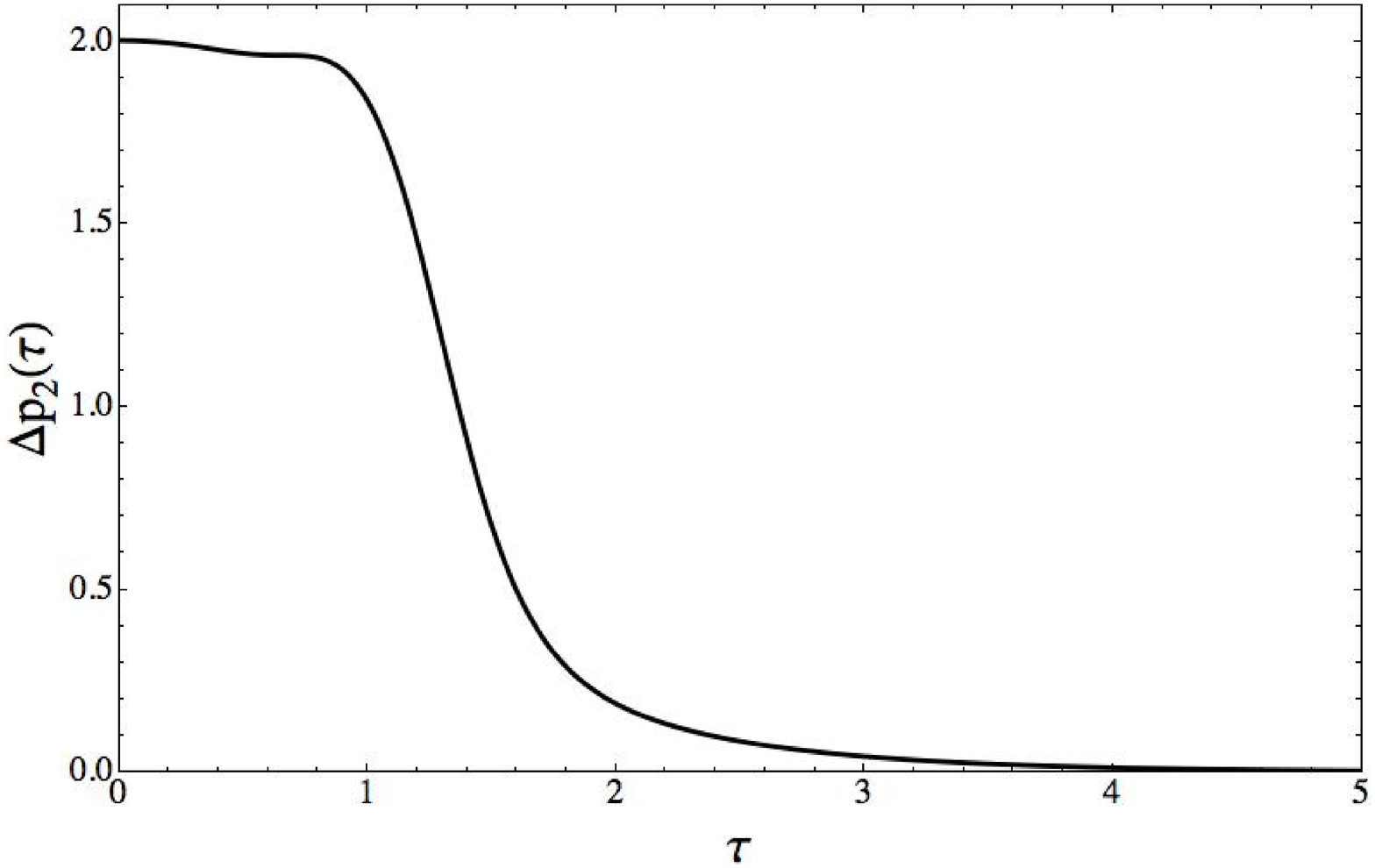}

    \caption{Relative difference in pressures for the first (left) and
second (right) profiles -- for $\tau \approx 1$ there is a rapid fall-off but
(perhaps, see text) does not reach yet a complete izotropization for the first
profile. }
    \end{center}
\end{figure}

\section{Summary}

This paper studies the early-times dynamics of boost-invariant plasma using
analytical methods. The motivations for this work are both 
phenomenological and theoretical. On the one hand,  the recent findings of the
RHIC experiment suggested that locally equilibrated 
nuclear matter behaves as an almost perfect fluid, which presumably indicates a
strongly coupled regime (at least at this stage of its 
evolution) of the underlying gauge field theory, $i.e.$ QCD. Despite  the fact
that realistic dynamics of the QCD after the collision 
would require the understanding of rapidity dependence ($i.e.$ deviations from 
boost-invariant dynamics, which is only approximately valid in the central 
rapidity region) and might be driven as well by the perturbative effects (as for
the Color Glass Condensate \cite{CGC} initial 
conditions), it is still interesting to consider the boost-invariant expansion
of a conformal plasma at strong coupling as a useful 
toy-model for $e.g.$  estimates of the thermalization time. On the other hand,
theoretically,  the boost-invariant plasma evolution is a 
workable example of a dynamical system at strong coupling, which is interesting
on its own. The modern developments translates (using 
the AdS/CFT correspondence) the dynamics of the strongly coupled gauge theories
into the evolution of higher dimensional space-time 
equipped with a nontrivial metric. Thus, as suggested previously by various
authors (e.g. \cite{Nastase:2005rp}) the thermalization of 
the excited gauge theory matter should be dual to  black hole (or black brane)
formation (see \cite{Chesler:2008hg, 
Bhattacharyya:2009uu} for concrete realizations of this observation), which is
obtained in late time as the dual of the Bjorken 
hydrodynamical flow. This subject is very fresh and thus any work which may shed
light on this fascinating process is valuable.

\vspace{10 pt}

The main result of the studies presented here is that the boost-invariant
dynamics of a strongly coupled conformal plasma is  sensitive 
to the initial conditions. This contradicts the scaling hypothesis
\cite{Kovchegov:2007pq}, which, analogous to the 
late-time case \cite{US1}  would indicate some uniqueness of the early-times
solution. In fact the scaling does not occur due to a subtlety --
the {\it a priori} dominant scaling contributions vanish precisely in the limit
$s\to 0.$  (see subsection (\ref{naiveET}) for details). The correct physical
picture 
leads to a link between the early-time expansion of the energy density and the
initial profile of the bulk metric. We find a quite general 
result that at all times, including the initial one, a singularity of at least
one  metric coefficient should develop in the bulk. Hence, the 
requirement that this does not lead to real curvature singularities already at
initial time is a basic constraint on the possible  initial 
conditions. 
%Another one is provided by the positivity of the energy-momentum
%tensor, constraint to be realized at all times, near the 
%physical boundary, see Eq.\eqref{positivity}. 
The analysis of the possible
curvature singularities in the initial data at finite distance from the 
boundary fixed the early times power series for the energy density to contain
only even powers of proper time. We have shown that 
solving the nonlinear constraint equation in the Fefferman - Graham coordinates
leads to the conclusion that the initial data must 
contain a coordinate singularity in the bulk of AdS which may signal the
presence of a \emph{dynamical} horizon (we are making a distinction w.r.t. the event horizon which is a global notion) right from the start of the
evolution. This suggests that the process of thermalization in this
boost-invariant setting does \emph{not} amount to a formation of a black hole,
as was
suggested up till now, but rather amounts to a qualitative difference in the
behaviour of the horizon. We plan to investigate these issues in more detail in
future work using numerical methods.

Interesting further directions of study include numerical investigation of the
bulk evolution dual to the boost-invariant flow. This 
can be achieved using the methods presented in \cite{Chesler:2008hg} and may
provide precise numerical results covering both early, 
intermediate and late time regimes (see \cite{CheslerTBA} for an interesting study of boost invariant flow sourced by boundary metric perturbations). It is important to understand
more qualitatively the relation between the initial 
conditions in the bulk and the shape of the energy density as function of proper
time, which would give the more precise estimates on 
the thermalization time. Moreover the methods developed in this paper are well
suited to reconsider the problem of plasma 
isotropisation posed in \cite{Janik:2008tc} using analytical methods (note
that anisotropic energy-momentum tensor should reach the 
equilibrium exponentially fast, whereas hydrodynamic evolution leaves a
power-like tail). Finally the most interesting, yet highly 
non-trivial, extensions of the AdS/CFT program for the  dynamical  evolution of
a plasma are the studies of the initial conditions such 
as  shock waves collision using the dual gravity picture (see for instance
\cite{Grumiller:2008va, Albacete:2008vs} for some 
preliminary attempts). We hope that the general properties and their numerical
implementation we found can serve as a testing ground 
for the investigation on the proper initial conditions and evolution of the
plasma. The authors plan to address some of the mentioned 
issues in the nearest future.

\acknowledgments
The authors would like to thank Ofer Aharony and Paul Chesler for discussions. During the course of this work, MH has been
supported by \emph{Polish/French Research 
Programme Polonium}, \emph{Polish Ministry of Science and Information
Technologies} grant N N202 247135 (2008-2010), \emph{Foundation 
for Polish Science} award START and \emph{Helena Kogutowska Scientific
Scholarship} by Jagiellonian University. MH and RJ would like to  thank the
Institute for Theortical Physics of Saclay for hospitality. 
MH and RJ were supported by Polish science funds during 2009-2011 as a research
project (NN202 105136).
The participation of
RP and RJ in this  investigation was partly supported by  6 Program of European
Union ``Marie Curie Transfer of Knowledge'' Project: Correlations in Complex
Systems ``COCOS'' MTKD-CT-2004-517186. R.P. wants to thank the Insitute of
physics of the Jagiellonian university in Cracow for hospitality.


\begin{thebibliography}{999}

\bibitem{review}
  E.~V.~Shuryak,
  ``What RHIC experiments and theory tell us about properties of  quark-gluon
  plasma?,''
  Nucl.\ Phys.\ A {\bf 750}, 64 (2005).
  %%CITATION = HEP-PH 0405066;%%

\bibitem{hydro}
 P.~F.~Kolb and U.~W.~Heinz,
  ``Hydrodynamic description of ultrarelativistic heavy-ion collisions,''
  [arXiv: \href{http://www.arxiv.org/abs/nucl-th/0305084}{nucl-th/0305084}];\\
  %%CITATION = NUCL-TH 0305084;%%
  P.~Huovinen and P.~V.~Ruuskanen,
  ``Hydrodynamic models for heavy ion collisions,''
  [arXiv: \href{http://www.arxiv.org/abs/nucl-th/0605008}{nucl-th/0605008}].
  %%CITATION = NUCL-TH 0605008;%%

\bibitem{JY}
  J.~Y.~Ollitrault,
  ``Anisotropy As A Signature Of Transverse Collective Flow,''
  Phys.\ Rev.\  D {\bf 46}, 229 (1992);\\
  See also the recent reviews:\ 
  J.~Y.~Ollitrault,
  ``Relativistic hydrodynamics,''
  Eur.\ J.\ Phys.\  {\bf 29}, 275 (2008);\\
  T.~Hirano, N.~van der Kolk and A.~Bilandzic,
  ``Hydrodynamic and Flow''.


\bibitem{adscft}    J.~M.~Maldacena,
  ``The large N limit of superconformal field theories and supergravity,''
  Adv.\ Theor.\ Math.\ Phys.\  {\bf 2}, 231 (1998)
  [Int.\ J.\ Theor.\ Phys.\  {\bf 38}, 1113 (1999)];\\
  %%CITATION = HEP-TH 9711200;%%
  S.~S.~Gubser, I.~R.~Klebanov and A.~M.~Polyakov,
  ``Gauge theory correlators from non-critical string theory,''
  Phys.\ Lett.\ B {\bf 428}, 105 (1998);\\
  %%CITATION = HEP-TH 9802109;%%
 E.~Witten,
  ``Anti-de Sitter space and holography,''
  Adv.\ Theor.\ Math.\ Phys.\  {\bf 2}, 253 (1998).
  %%CITATION = HEP-TH 9802150;%%
  
%\cite{Kovtun:2004de}
\bibitem{son}
  G.~Policastro, D.~T.~Son and A.~O.~Starinets,
  ``The shear viscosity of strongly coupled N = 4 supersymmetric Yang-Mills
  plasma,''
  Phys.\ Rev.\ Lett.\  {\bf 87}, 081601 (2001);\\
P.~Kovtun, D.~T.~Son and A.~O.~Starinets,
  ``Viscosity in strongly interacting quantum field theories from black hole
  physics,''
  Phys.\ Rev.\ Lett.\  {\bf 94}, 111601 (2005).
  %%CITATION = PRLTA,94,111601;%%

\bibitem{Bjorken}
 J.~D.~Bjorken,
  ``Highly Relativistic Nucleus-Nucleus Collisions: The Central Rapidity
  Region,''
  Phys.\ Rev.\ D {\bf 27}, 140 (1983).
  %%CITATION = PHRVA,D27,140;%%


\bibitem{US1}
  R.~A.~Janik and R.~B.~Peschanski,
  ``Asymptotic perfect fluid dynamics as a consequence of AdS/CFT,''
  Phys.\ Rev.\  D {\bf 73}, 045013 (2006),\\
  ``Gauge / gravity duality and thermalization of a boost-invariant perfect
  fluid,''
  Phys.\ Rev.\  D {\bf 74}, 046007 (2006).
  
 
 %\cite{MINWALLA}
\bibitem{MINWALLA}
  S.~Bhattacharyya, V.~E.~Hubeny, S.~Minwalla and M.~Rangamani,
  ``Nonlinear Fluid Dynamics from Gravity,''
  JHEP {\bf 0802}, 045 (2008).
  %%CITATION = JHEPA,0802,045;%%

\bibitem{CGC}
  For a review: E.~Iancu and R.~Venugopalan,
  ``The color glass condensate and high energy scattering in QCD,''
  arXiv:hep-ph/0303204.
  %%CITATION = HEP-PH/0303204;%%

\bibitem{beuf}
  G.~Beuf,
  ``Gravity dual of N=4 SYM theory with fast moving sources,''
  arXiv:0903.1047 [hep-th].
  %%CITATION = ARXIV:0903.1047;%%

%\cite{RJ}
\bibitem{RJ}
  R.~A.~Janik,
  ``Viscous plasma evolution from gravity using AdS/CFT,''
  Phys.\ Rev.\ Lett.\  {\bf 98}, 022302 (2007).
  %%CITATION = PRLTA,98,022302;%%
  
%\cite{Heller:2007qt}
\bibitem{Heller:2007qt}
  M.~P.~Heller and R.~A.~Janik,
  ``Viscous hydrodynamics relaxation time from AdS/CFT,''
  Phys.\ Rev.\  D {\bf 76}, 025027 (2007).
  %%CITATION = PHRVA,D76,025027;%%
%\cite{Visser:1994jb}
  
  
\bibitem{REF}
  H.~Epstein, V.~Glaser and A.~Jaffe,
  ``Nonpositivity of energy density in Quantized field theories,''
  Nuovo Cim.\  {\bf 36}, 1016 (1965), quoted in Ref.\cite{Grumiller:2008va}.
  %%CITATION = JHEPA,0808,027;%%



\bibitem{fg} C. Fefferman and C.R. Graham, ``Conformal Invariants,''
  in {\it Elie Cartan et les Math\'ematiques d'aujourd'hui},
  Ast\'erisque (1985) 95.
  
  \bibitem{Skenderis}
S.~de Haro, S.~N.~Solodukhin and K.~Skenderis,
  ``Holographic reconstruction of spacetime and renormalization in the
  AdS/CFT correspondence,''
  Commun.\ Math.\ Phys.\  {\bf 217}, 595 (2001)
  [arXiv: \href{http://www.arxiv.org/abs/hep-th/0002230}{hep-th/0002230}];\\
  %%CITATION = HEP-TH 0002230;%%
 K.~Skenderis,
  ``Lecture notes on holographic renormalization,''
  Class.\ Quant.\ Grav.\  {\bf 19}, 5849 (2002)
  [arXiv: \href{http://www.arxiv.org/abs/hep-th/0209067}{hep-th/0209067}].
  %%CITATION = HEP-TH 0209067;%%
  
 
%\cite{Nak1}
\bibitem{Nak1}
  S.~Nakamura and S.~J.~Sin,
  ``A holographic dual of hydrodynamics,''
  JHEP {\bf 0609}, 020 (2006).
  %%CITATION = HEP-TH 0607123;%%
  
  
\bibitem{Bak:2006dn}
  D.~Bak and R.~A.~Janik,
  ``From static to evolving geometries: R-charged hydrodynamics from supergravity,''
  Phys.\ Lett.\  B {\bf 645}, 303 (2007).
  %%CITATION = PHLTA,B645,303;%%
  
  
%\cite{Benincasa:2007tp}
\bibitem{Benincasa:2007tp}
  P.~Benincasa, A.~Buchel, M.~P.~Heller and R.~A.~Janik,
  ``On the supergravity description of boost invariant conformal plasma at
  strong coupling,''
  Phys.\ Rev.\  D {\bf 77}, 046006 (2008).
  %%CITATION = PHRVA,D77,046006;%%
  
%\cite{Buchel:2008xr}
\bibitem{Buchel:2008xr}
  A.~Buchel,
  ``On SUGRA description of boost-invariant conformal plasma at strong coupling,''
  AIP Conf.\ Proc.\  {\bf 1031}, 196 (2008)
  [arXiv:0803.3421 [hep-th]].
  %%CITATION = APCPC,1031,196;%%
  

%\cite{Heller:2008mb}
\bibitem{Heller:2008mb}
  M.~P.~Heller, P.~Surowka, R.~Loganayagam, M.~Spalinski and S.~E.~Vazquez,
  ``On a consistent AdS/CFT description of boost-invariant plasma,'' Phys.\ Rev.\ Lett.\  {\bf 102}, 041601 (2009),
  arXiv:0805.3774 [hep-th].
  %%CITATION = ARXIV:0805.3774;%%
  
  
  
%\cite{Kinoshita:2008dq}
\bibitem{Kinoshita:2008dq}
  S.~Kinoshita, S.~Mukohyama, S.~Nakamura and K.~y.~Oda,
  ``A Holographic Dual of Bjorken Flow,''
  arXiv:0807.3797 [hep-th].
  %%CITATION = ARXIV:0807.3797;%%

 
  
%\cite{Heller:2008fg}
\bibitem{Heller:2008fg}
  M.~P.~Heller, R.~A.~Janik and R.~Peschanski,
  ``Hydrodynamic Flow of the Quark-Gluon Plasma and Gauge/Gravity Correspondence,''
  Acta Phys.\ Polon.\  B {\bf 39}, 3183 (2008)
  [arXiv:0811.3113 [hep-th]].
  %%CITATION = APPOA,B39,3183;%%
  

  
 %\cite{Kovchegov:2007pq}
\bibitem{Kovchegov:2007pq}
  Y.~V.~Kovchegov and A.~Taliotis,
  ``Early time dynamics in heavy ion collisions from AdS/CFT correspondence,''
  Phys.\ Rev.\  C {\bf 76}, 014905 (2007).
  %%CITATION = PHRVA,C76,014905;%%

%\cite{Figueras:2009iu}
\bibitem{Figueras:2009iu}
  P.~Figueras, V.~E.~Hubeny, M.~Rangamani and S.~F.~Ross,
  ``Dynamical black holes and expanding plasmas,''
  arXiv:0902.4696 [hep-th].
  %%CITATION = ARXIV:0902.4696;%%
  
 
 
  %\cite{Janik:2008tc}
\bibitem{Janik:2008tc}
  R.~A.~Janik and P.~Witaszczyk,
  ``Towards the description of anisotropic plasma at strong coupling,''
  JHEP {\bf 0809}, 026 (2008).
  %%CITATION = JHEPA,0809,026;%% %%CITATION = ARXIV:0809.5033;%%

  %\cite{Buchel:2000ch}
\bibitem{Buchel:2000ch}
  A.~Buchel,
  ``Finite temperature resolution of the Klebanov-Tseytlin singularity,''
  Nucl.\ Phys.\  B {\bf 600}, 219 (2001)
  [arXiv:hep-th/0011146].
  %%CITATION = NUPHA,B600,219;%%
  
  
%\cite{Baier:2007ix}
\bibitem{Baier:2007ix}
  R.~Baier, P.~Romatschke, D.~T.~Son, A.~O.~Starinets and M.~A.~Stephanov,
  ``Relativistic viscous hydrodynamics, conformal invariance, and holography,''
  JHEP {\bf 0804}, 100 (2008).
  %%CITATION = JHEPA,0804,1
  

\bibitem{BoothHellerSpalinski}
  I.~Booth, M.~P.~Heller and M.~Spalinski,
  ``Black brane entropy and hydrodynamics: the boost-invariant case,''
  arXiv:0910.0748 [hep-th].
  %%CITATION = ARXIV:0910.0748;%%


%\cite{Booth:2005qc}
\bibitem{Booth:2005qc}
  I.~Booth,
  ``Black hole boundaries,''
  Can.\ J.\ Phys.\  {\bf 83}, 1073 (2005)
  [arXiv:gr-qc/0508107].
  %%CITATION = CJPHA,83,1073;%%

 %\cite{Nastase:2005rp}
\bibitem{Nastase:2005rp}
  H.~Nastase,
  ``The RHIC fireball as a dual black hole,''
  arXiv:hep-th/0501068.
  %%CITATION = HEP-TH/0501068;%%


%\cite{Chesler:2008hg}
\bibitem{Chesler:2008hg}
  P.~M.~Chesler and L.~G.~Yaffe,
  ``Horizon formation and far-from-equilibrium isotropization in supersymmetric
  Yang-Mills plasma,''
  arXiv:0812.2053 [hep-th].
  %%CITATION = ARXIV:0812.2053;%%
  
%\cite{Bhattacharyya:2009uu}
\bibitem{Bhattacharyya:2009uu}
  S.~Bhattacharyya and S.~Minwalla,
  ``Weak Field Black Hole Formation in Asymptotically AdS Spacetimes,''
  JHEP {\bf 0909}, 034 (2009)
  [arXiv:0904.0464 [hep-th]].
  %%CITATION = JHEPA,0909,034;%%
  
%\cite{CheslerTBA}
\bibitem{CheslerTBA}
P.~M.~Chesler and L.~G.~Yaffe,
  ``Boost invariant flow, black hole formation, and far-from-equilibrium dynamics in \boldmath $\mathcal N = 4$~supersymmetric Yang-Mills theory,''
  arXiv:0906.4426 [hep-th].
  %%CITATION = ARXIV:0906.4426;%%


%\cite{Grumiller:2008va}
\bibitem{Grumiller:2008va}
  D.~Grumiller and P.~Romatschke,
  ``On the collision of two shock waves in AdS5,''
  JHEP {\bf 0808}, 027 (2008).
  %%CITATION = JHEPA,0808,027;%%

%\cite{Albacete:2008vs}
\bibitem{Albacete:2008vs}
  J.~L.~Albacete, Y.~V.~Kovchegov and A.~Taliotis,
  ``Modeling Heavy Ion Collisions in AdS/CFT,''
  JHEP {\bf 0807}, 100 (2008).
  %%CITATION = JHEPA,0807,100;%%
  

\end{thebibliography}
\end{document}